\documentclass{article}
\pdfoutput=1

\usepackage[utf8]{inputenc}
\usepackage{amsmath, amsthm, amssymb, amsfonts, dsfont, mathrsfs, dirtytalk, tikz-cd, adjustbox, url, nicefrac, booktabs,array, graphicx,todonotes,longtable, tabu,bm,cite}
\usepackage[T1]{fontenc}    
\usepackage[english]{babel}
\usepackage{hyperref}
\usepackage[capitalise]{cleveref}
\usepackage{arxiv}

\newcommand{\E}{\mathbb E}

\newtheorem{theorem}{Theorem}
\newtheorem{lemma}[theorem]{Lemma}

\newtheorem{corollary}[theorem]{Corollary}
\newtheorem{remark}[theorem]{Remark}

\title{Active inference on discrete state-spaces -- a synthesis}
\author{Lancelot Da Costa$^{1,2,}$\thanks{Author correspondence: \texttt{l.da-costa@imperial.ac.uk}}\:,\: Thomas Parr$^2$, \:Noor Sajid$^2$, \: Sebastijan Veselic$^2$,\: Victorita Neacsu$^2$,\: Karl Friston$^2$ \\ \\
	$^1$Department of Mathematics, Imperial College London \\ \\
	$^2$Wellcome Centre for Human Neuroimaging, University College London}
\begin{document}
	
	\maketitle
	
	\begin{abstract}
		Active inference is a normative principle underwriting perception, action, planning, decision-making and learning in biological or artificial agents. From its inception, its associated process theory has grown to incorporate complex generative models, enabling simulation of a wide range of complex behaviours. Due to successive developments in active inference, it is often difficult to see how its underlying principle relates to process theories and practical implementation. In this paper, we try to bridge this gap by providing a complete mathematical synthesis of active inference on discrete state-space models. This technical summary provides an overview of the theory, derives neuronal dynamics from first principles and relates this dynamics to biological processes. Furthermore, this paper provides a fundamental building block needed to understand active inference for mixed generative models; allowing continuous sensations to inform discrete representations. This paper may be used as follows: to guide research towards outstanding challenges, a practical guide on how to implement active inference to simulate experimental behaviour, or a pointer towards various in-silico neurophysiological responses that may be used to make empirical predictions.
	\end{abstract}
	
	\textbf{Keywords}: active inference, free energy principle, process theory, variational Bayesian inference, Markov decision process, explained.
	
	\tableofcontents

	\section{Introduction}
	Active inference is a normative principle underlying perception, action, planning, decision-making and learning in biological or artificial agents. It postulates that these processes may all be seen as optimising two complementary objective functions; namely, a variational free energy, which measures the fit between an internal model and (past) sensory observations, and an expected free energy, which scores possible (future) courses of action in relation to prior preferences. Active inference has been employed to simulate a wide range of complex behaviours, including planning and navigation \cite{kaplanPlanningNavigationActive2018}, reading \cite{fristonDeepTemporalModels2018}, curiosity and abstract rule learning \cite{fristonActiveInferenceCuriosity2017}, saccadic eye movements \cite{parrActiveInferenceAnatomy2018}, visual foraging \cite{mirzaSceneConstructionVisual2016,parrUncertaintyEpistemicsActive2017}, visual neglect \cite{parrComputationalAnatomyVisual2018}, hallucinations \cite{adamsComputationalAnatomyPsychosis2013}, niche construction \cite{bruinebergFreeenergyMinimizationJoint2018,constantVariationalApproachNiche2018}, social conformity \cite{constantRegimesExpectationsActive2019}, impulsivity \cite{mirzaImpulsivityActiveInference2019}, image recognition \cite{millidgeImplementingPredictiveProcessing2019}, and the mountain car problem \cite{catalBayesianPolicySelection2019,fristonReinforcementLearningActive2009,fristonWhatValueAccumulated2012}. The key idea that underwrites these simulations is that creatures use an internal forward (generative) model to predict their sensory input, which they use to infer the causes of these data.
	
	Early formulations of active inference employed generative models expressed in continuous space and time (for an introduction see \cite{bogaczTutorialFreeenergyFramework2017}, for a review see \cite{buckleyFreeEnergyPrinciple2017}), with behaviour modelled as a continuously evolving random dynamical system. However, we know that some processes in the brain conform better to discrete, hierarchical, representations, compared to continuous representations (e.g., visual working memory \cite{luckCapacityVisualWorking1997, zhangDiscreteFixedResolutionRepresentations2008}, state estimation via place cells \cite{eichenbaumHippocampusMemoryPlace1999, okeefeHippocampusSpatialMap1971}, language, etc). Reflecting this, many of the paradigms studied in neuroscience are naturally framed as discrete state-space problems. Decision-making tasks are a prime candidate for this, as they often entail a series of discrete alternatives that an agent needs to choose among (e.g., multi-arm bandit tasks \cite{dawCorticalSubstratesExploratory2006, reverdyModelingHumanDecisionmaking2013, wuGeneralizationGuideshuman2018}, multi-step decision tasks \cite{dawModelBasedInfluencesHumans2011}). This explains why – in active inference – agent behaviour is often modelled using a discrete state-space formulation: the particular applications of which are summarised in Table \ref{table:1}. More recently, mixed generative models \cite{fristonGraphicalBrainBelief2017} – combining discrete and continuous states – have been used to model behaviour involving discrete and continuous representations (e.g., decision-making and movement \cite{parrDiscreteContinuousBrain2018}, speech production and recognition \cite{fristonActiveListening2020}, pharmacologically induced changes in eye-movement control \cite{parrComputationalPharmacologyOculomotion2019} or reading; involving continuous visual sampling informing inferences about discrete semantics \cite{fristonGraphicalBrainBelief2017}).
	
	\begin{longtabu} to \textwidth {
			X[2,c]
			X[4,c]
			X[1,c]}
		\caption{Applications of active inference (discrete state-space).} \label{table:1} \\
		\toprule
		Application & Description & References \\
		\midrule
		Decision-making under uncertainty
		& Initial formulation of active inference on partially observable Markov decision processes.
		& \cite{fristonActiveInferenceAgency2012}  \\ \addlinespace[0.3cm]
		Optimal control
		&  {Application of KL or risk sensitive control in an engineering benchmark – the mountain car problem.}
		&  {\cite{fristonWhatValueAccumulated2012,catalBayesianPolicySelection2019}} \\ \addlinespace[0.3cm]
		{Evidence accumulation}
		&	 {Illustrating the role of evidence accumulation in decision-making through an urns task.}
		&	 {\cite{fitzgeraldActiveInferenceEvidence2015, fitzgeraldPrecisionNeuronalDynamics2015} }\\ \addlinespace[0.3cm]
		{Psychopathology}
		&  {Simulation of addictive choice behaviour.}
		&  {\cite{schwartenbeckOptimalInferenceSuboptimal2015}} \\ \addlinespace[0.3cm]
		{Dopamine}
		&  {The precision of beliefs about policies provides a plausible description of dopaminergic discharges.}
		&  {\cite{fitzgeraldDopamineRewardLearning2015,fristonAnatomyChoiceDopamine2014}} \\ \addlinespace[0.3cm]
		{Functional magnetic resonance imaging}
		&	 {Empirical prediction and validation of dopaminergic discharges.}
		&  {\cite{schwartenbeckDopaminergicMidbrainEncodes2015}}\\ \addlinespace[0.3cm]
		{Maximal utility theory}
		&  {Evidence in favour of surprise minimization as opposed to utility maximisation in human decision-making.}
		&  {\cite{schwartenbeckEvidenceSurpriseMinimization2015}}\\ \addlinespace[0.3cm]
		{Social cognition}
		&  {Examining the effect of prior preferences on interpersonal inference.}
		&  {\cite{moutoussisFormalModelInterpersonal2014}}\\ \addlinespace[0.3cm]
		{Exploration-exploitation dilemma} &  {Casting behaviour as expected free energy minimising accounts for epistemic and pragmatic choices.} &  {\cite{fristonActiveInferenceEpistemic2015}}\\ \addlinespace[0.3cm]
		{Habit learning and action selection}&  {Formulating learning as an inferential process and action selection as Bayesian model averaging.} &  {\cite{fristonActiveInferenceLearning2016,fitzgeraldModelAveragingOptimal2014}}\\ \addlinespace[0.3cm]
		{Scene construction and anatomy of time} &  {Mean-field approximation for multi-factorial hidden states, enabling high dimensional representations of the environment.} &  {\cite{fristonFunctionalAnatomyTime2016,mirzaSceneConstructionVisual2016}}\\ \addlinespace[0.3cm]
		{Electrophysiological responses}  &	 {Synthesising various in-silico neurophysiological responses via a gradient descent on free energy. E.g., place-cell activity, mismatch negativity, phase-precession, theta sequences, theta-gamma coupling and dopaminergic discharges. }	&   {\cite{fristonActiveInferenceProcess2017}}\\\addlinespace[0.3cm]
		{Structure learning, curiosity and insight} &	 {Simulation of artificial curiosity and abstract rule learning. Structure learning via Bayesian model reduction.} &  {\cite{fristonActiveInferenceCuriosity2017}}\\\addlinespace[0.3cm]
		{Hierarchical temporal representations }&	 {Generalisation to hierarchical generative models with deep temporal structure and simulation of reading.} &  {\cite{fristonDeepTemporalModels2018,parrWorkingMemoryAttention2017}}\\\addlinespace[0.3cm]
		{Computational neuropsychology} &  {Simulation of visual neglect, hallucinations, and prefrontal syndromes under alternative pathological priors. }&  {\cite{parrComputationalAnatomyVisual2018,parrComputationalNeuropsychologyBayesian2018,parrPrefrontalComputationActive2019,benrimohActiveInferenceAuditory2018,parrPrecisionFalsePerceptual2018}}\\\addlinespace[0.3cm]
		{Neuromodulation}	&  {Use of precision parameters to manipulate exploration during saccadic searches; associating uncertainty with cholinergic and noradrenergic systems.} &  {\cite{parrUncertaintyEpistemicsActive2017,parrComputationalPharmacologyOculomotion2019,salesLocusCoeruleusTracking2018,vincentEyeUncertaintyModelling2019}} \\\addlinespace[0.3cm]
		{Decisions to movements} &  {Mixed generative models combining discrete and continuous states to implement decisions through movement.} &  {\cite{fristonGraphicalBrainBelief2017,parrDiscreteContinuousBrain2018}} \\\addlinespace[0.3cm]
		{Planning, navigation and niche construction	}&  {Agent induced changes in environment (generative process); decomposition of goals into subgoals.} &  {\cite{kaplanPlanningNavigationActive2018,bruinebergFreeenergyMinimizationJoint2018,constantVariationalApproachNiche2018}}\\\addlinespace[0.3cm]
		{Atari games}	&  {Active inference compares favourably to reinforcement learning in the game of Doom.} &  {\cite{cullenActiveInferenceOpenAI2018}}\\\addlinespace[0.3cm]
		{Machine learning} &  {Scaling active inference to more complex machine learning problems.} &  {\cite{tschantzScalingActiveInference2019}} \\ \addlinespace[0.15cm]
		\bottomrule
	\end{longtabu}
	
	Due to the pace of recent theoretical advances in active inference, it is often difficult to retain a comprehensive overview of its process theory and practical implementation. In this paper, we hope to provide a comprehensive (mathematical) synthesis of active inference on discrete state-space models. This technical summary provides an overview of the theory, derives the associated (neuronal) dynamics from first principles and relates these to known biological processes. Furthermore, this paper and \cite{buckleyFreeEnergyPrinciple2017} provide the building blocks necessary to understand active inference on mixed generative models. This paper can be read as a practical guide on how to implement active inference for simulating experimental behaviour, or a pointer towards various in-silico neuro- and electro- physiological responses that can be tested empirically.
	
	This paper is structured as follows. Section \ref{sec: active inference} is a high-level overview of active inference. The following sections elucidate the formulation by deriving the entire process theory from first principles; incorporating perception, planning and decision-making. This formalises the action-perception cycle: 1) an agent is presented with a stimulus, 2) it infers its latent causes, 3) plans into the future and 4) realises its preferred course of action; and repeat. This enactive cycle allows us to explore the dynamics of synaptic plasticity, which mediate learning of the contingencies of the world at slower timescales. We conclude in section \ref{sec: structure learning} with an overview of structure learning in active inference.
	
	\section{Active inference}
	\label{sec: active inference}
	
	To survive in a changing environment, biological (and artificial) agents must maintain their sensations within a certain hospitable range (i.e., maintaining homeostasis through allostasis). In brief, active inference proposes that agents achieve this by optimising two complementary objective functions, a variational free energy and an expected free energy. In short, the former measures the fit between an internal (generative) model of its sensations and sensory observations, while the latter scores each possible course of action in terms of its ability to reach the range of “preferred” states of being.
	
	Our first premise is that agents represent the world through an internal model. Through minimisation of variational free energy, this model becomes a good model of the environment. In other words, this probabilistic model and the probabilistic beliefs\footnote{By beliefs we mean Bayesian beliefs, i.e., probability distributions over a variable of interest (e.g., current position). Beliefs are therefore used in the sense of Bayesian belief updating or belief propagation – as opposed to propositional or folk psychology beliefs.} that it encodes are continuously updated to mirror the environment and its dynamics. Such a world model is considered to be generative; in that it is able to generate predictions about sensations (e.g., during planning or dreaming), given beliefs about future states of being. If an agent senses a heat source (e.g., another agent) via some temperature receptors, the sensation of warmth represents an observed outcome and the temperature of the heat source a hidden state; minimisation of variational free energy then ensures that beliefs about hidden states closely match the true temperature. Formally, the generative model is a joint probability distribution over possible hidden states and sensory consequences – that specifies how the former cause the latter -- and minimisation of variational free energy enables to "invert" the model; i.e., determine the most likely hidden states given sensations. The variational free energy is the negative evidence lower bound that is optimised in variational Bayes in machine learning \cite{bishopPatternRecognitionMachine2006,xitongUnderstandingVariationalLower2017}. Technically – by minimising variational free energy – agents perform approximate Bayesian inference \cite{senguptaApproximateBayesianInference2016,senguptaNeuronalGaugeTheory2016}, which enables them to infer the causes of their sensations (e.g., perception). This is the point of contact between active inference and the Bayesian brain \cite{aitchisonYouPredictiveCoding2017,fristonHistoryFutureBayesian2012,knillBayesianBrainRole2004}. Crucially, agents may incorporate an optimism bias \cite{mckayEvolutionMisbelief2009,sharotOptimismBias2011} in their model; thereby scoring certain “preferred” sensations as more likely. This lends a higher plausibility to those courses of action that realise these sensations. In other words, a preference is simply something an agent (believes it) is likely to work towards.
	
	To maintain homeostasis, and ensure survival, agents must minimise surprise\footnote{In information theory, the surprise (a.k.a., surprisal) associated with an outcome under a generative model is given by $-\log p(o)$. This specifies the extent to which an observation is unusual and surprises the agent – but this does not mean that the agent consciously experiences surprise. In information theory this kind of surprise is known as self-information.}. Since the generative model scores preferred outcomes as more likely, minimising surprise corresponds to maximising model evidence\footnote{In Bayesian statistics, the model evidence (often referred to as marginal likelihood) associated with a generative model is $p(o)$ -- the probability of observed outcomes according to the model (sometimes this is written as $p(o|m)$, explicitly conditioning upon a model). The model evidence scores the goodness of the model as an explanation of data that are sampled, by rewarding accuracy and penalising complexity, which avoids overfitting.}). In active inference, this is assured by the aforementioned processes; indeed, the variational free energy turns out to be an upper bound on surprise and minimising expected free energy ensures preferred outcomes are realised, thereby avoiding surprise on average.
	
	Active inference can thus be framed as the minimisation of surprise \cite{fristonFreeenergyPrincipleRough2009,fristonFreeenergyPrincipleUnified2010,fristonFreeenergyBrain2007,fristonFreeEnergyPrinciple2006} by perception and action. In discrete state models -- of the sort discussed here -- this means agents select from different possible courses of action (i.e., policies) in order to realise their preferences and thus minimise the surprise that they expect to encounter in the future. This enables a Bayesian formulation of the perception-action cycle \cite{fusterPrefrontalCortexBridging1990}: agents perceive the world by minimising variational free energy, ensuring their model is consistent with past observations, and act by minimising expected free energy, to make future sensations consistent with their model. This account of behaviour can be concisely framed as self-evidencing \cite{hohwySelfEvidencingBrain2016}.
	
\begin{figure}
		\centering
		\includegraphics[width=\textwidth]{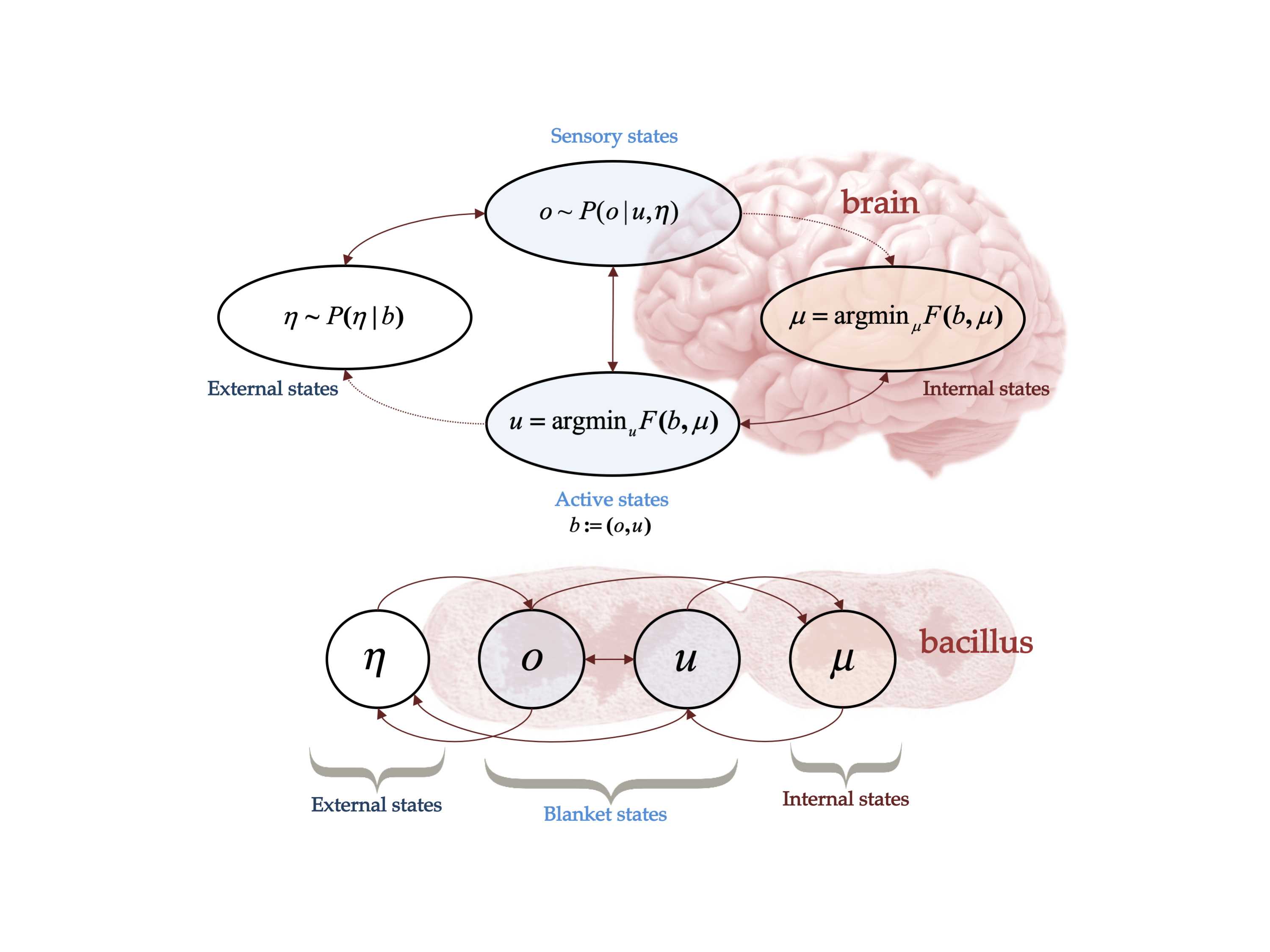}
		\caption{\textbf{Markov blankets in active inference.} This figure illustrates the Markov blanket assumption of active inference. A Markov blanket is a set of variables through which states internal and external to the system interact. Specifically, the system must be such that we can partition it into a Bayesian network of internal states $\mu$, external states $\eta$, sensory states $o$ and active states $u$, ($\mu$, $o$ and $u$ are often referred together as \textit{particular states}) with probabilistic (causal) links in the directions specified by the arrows. All interactions between internal and external states are therefore mediated by the blanket states $b$. The sensory states represent the sensory information that the body receives from the environment and the active states express how the body influences the environment. This blanket assumption is quite generic, in that it can be reasonably assumed for a brain as well as elementary organisms. For example, when considering a bacillus, the sensory states become the cell membrane and the active states comprise the actin filaments of the cytoskeleton. Under the Markov blanket assumption – together with the assumption that the system persists over time (i.e., possesses a non-equilibrium steady state) – a generalised synchrony appears, such that the dynamics of the internal states can be cast as performing inference over the external states (and vice-versa) via a minimisation of variational free energy \cite{parrMarkovBlanketsInformation2019,fristonFreeEnergyPrinciple2019}. This coincides with existing approaches to inference; i.e., variational Bayes \cite{bishopPatternRecognitionMachine2006,bealVariationalAlgorithmsApproximate2003,bleiVariationalInferenceReview2017,jordanIntroductionVariationalMethods1998}. This can be viewed as the internal states mirroring external states, via sensory states (e.g., perception), and external states mirroring internal states via active states (e.g., a generalised form of self-assembly, autopoiesis or niche construction). Furthermore, under these assumptions the most likely courses of actions can be shown to minimise expected free energy. Note that external states beyond the system should not be confused with the hidden states of the agent’s generative model (which model external states). In fact, the internal states are exactly the parameters (i.e., sufficient statistics) encoding beliefs about hidden states and other latent variables, which model external states in a process of variational free energy minimisation. Hidden and external states may or may not be isomorphic. In other words, an agent uses its internal states to represent hidden states that may or may not exist in the external world.}
		\label{fig: markov blanket}
	\end{figure}

	In contrast to other normative models of behaviour, active inference is a ‘first principle’ account, which is grounded in statistical physics \cite{parrMarkovBlanketsInformation2019,fristonFreeEnergyPrinciple2019}. Active inference describes the dynamics of systems that persist (i.e., do not dissipate) during some timescale of interest, and that can be statistically segregated from their environment -- conditions which are satisfied by biological systems. Mathematically, the first condition means that the system is at \textit{non-equilibrium steady-state} (NESS). This implies the existence of a steady-state probability density to which the system self-organises and returns to after perturbation (i.e., the agent's preferences). The statistical segregation condition is the presence of a Markov blanket (c.f., Figure \ref{fig: markov blanket}) \cite{kirchhoffMarkovBlanketsLife2018,pearlGraphicalModelsProbabilistic1998}: a set of variables through which states internal and external to the system interact (e.g., the skin is a Markov blanket for the human body). Under these assumptions it can be shown that the states internal to the system parameterise Bayesian beliefs about external states and can be cast a process of variational free energy minimisation. This coincides with existing approaches to approximate inference \cite{bishopPatternRecognitionMachine2006,bealVariationalAlgorithmsApproximate2003,bleiVariationalInferenceReview2017,jordanIntroductionVariationalMethods1998}. Furthermore, it can be shown that the most likely courses of action taken by those systems are those which minimise expected free energy -- a quantity that subsumes many existing constructs in science and engineering (see section \ref{sec: efe}).
	
	By subscribing to the above assumptions, it is possible to describe the behaviour of viable living systems as performing active inference -- the remaining challenge is to determine the computational and physiological processes that they implement to do so. This paper aims to summarise a possible answers to this question, by reviewing the technical details of a process theory for active inference on discrete state-space generative models, first presented in \cite{fristonActiveInferenceProcess2017}. Note that it is important to distinguish between active inference as a principle (presented above) from active inference as a process theory. The former is a consequence of fundamental assumptions about living systems, while the latter is a hypothesis concerning the computational and biological processes in the brain that might implement active inference. The ensuing process theories theory can then be used to predict plausible neuronal dynamics and electrophysiological responses that are elicited experimentally.
	
	\section{Discrete state-space generative models}
	
\begin{figure}
		\centering
		\includegraphics[width=\textwidth]{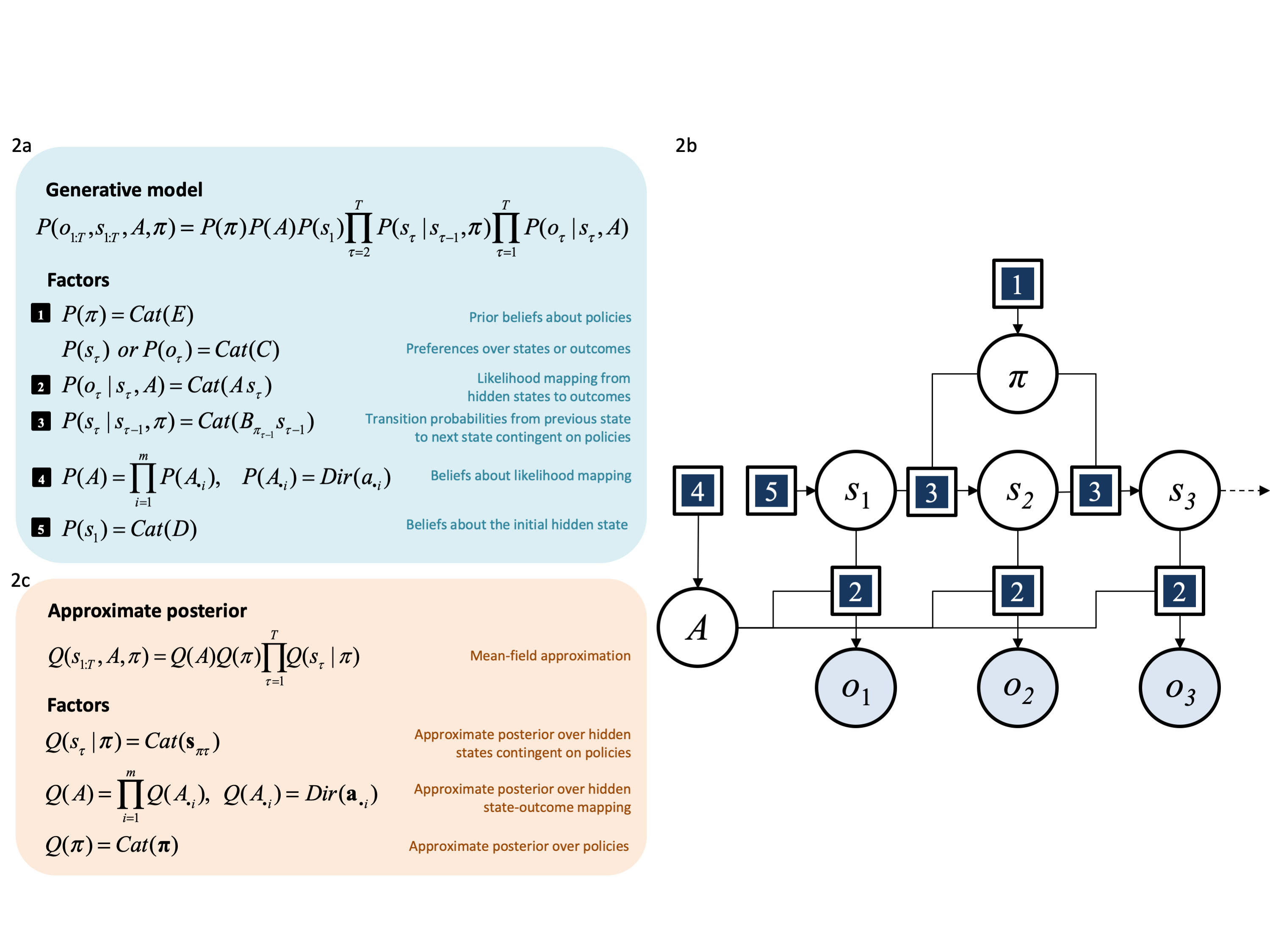}
		\caption{\textbf{Example of a discrete state-space generative model.} Panel 2a, specifies the form of the generative model, which is how the agent represents the world. The generative model is a joint probability distribution over (hidden) states, outcomes and other variables that cause outcomes. In this representation, states unfold in time causing an observation at each time-step. The likelihood matrix $A$ encodes the probabilities of state-outcome pairs. The policy $\pi$ specifies which action to perform at each time-step. Note that the agent's preferences may be specified either in terms of states or outcomes. It is important to distinguish between states (resp. outcomes) that are random variables, and the possible values that they can take in $S$ (resp. in $O$), which we refer to as possible states (resp. possible outcomes). Note that this type of representation comprises a finite number of timesteps, actions, policies, states, outcomes, possible states and possible outcomes. In Panel 2b, the generative model is displayed as a probabilistic graphical model \cite{bishopPatternRecognitionMachine2006,pearlGraphicalModelsProbabilistic1998,jordanIntroductionVariationalMethods1998,pearlProbabilisticReasoningIntelligent1988} expressed in factor graph form \cite{loeligerIntroductionFactorGraphs2004}. The variables in circles are random variables, while squares represent factors, whose specific form are given in Panel 2a. The arrows represent causal relationships (i.e., conditional probability distributions). The variables highlighted in grey can be observed by the agent, while the remaining variables are inferred through approximate Bayesian inference (see Section \ref{sec: VB}) and called hidden or latent variables. Active inference agents perform inference by optimising the parameters of an approximate posterior distribution (see Section \ref{sec: VB}). Panel 2c specifies how this approximate posterior factorises under a particular mean-field approximation \cite{tanakaTheoryMeanField1999}, although other factorisations may be used \cite{parrNeuronalMessagePassing2019,schwobelActiveInferenceBelief2018}. A glossary of terms used in this figure is available in Table \ref{table:2}. The mathematical yoga of generative models is heavily dependent on Markov blankets. The Markov blanket of a random variable in a probabilistic graphical model are those variables that share a common factor. Crucially, a variable conditioned upon its Markov blanket is conditionally independent of all other variables. We will use this property extensively (and implicitly) in the text.}
		\label{fig:gen mod}
	\end{figure}

	The generative model \cite{bishopPatternRecognitionMachine2006} expresses how the agent represents the world. This is a joint probability distribution over sensory data and the hidden (or latent) causes of these data. The sorts of discrete state-space generative models used in active inference are specifically suited to represent discrete time series and decision-making tasks. These can be expressed as variants of partially observable Markov decision processes (POMDPs; \cite{astromOptimalControlMarkov1965}): from simple Markov decision processes \cite{bartoReinforcementLearningIntroduction1992,stoneArtificialIntelligenceEngines2019,whiteMarkovDecisionProcesses2001} to generalisations in the form of deep probabilistic (hierarchical) models \cite{fristonDeepTemporalModels2018,boxMultiparameterProblemsBayesian1965,allenbyHierarchicalBayesModels2005}. For clarity, the process theory is derived for the simplest model that facilitates understanding of subsequent generalisations; namely, a POMDP where the agent holds beliefs about the probability of the initial state (specified as $D$), the transition probabilities from one state to the next (defined as matrix $B$) and the probability of states given outcomes (i.e., the likelihood matrix $A$); see Figure \ref{fig:gen mod}.
	
	\begin{longtabu} to \textwidth {
			X[1,c]
			X[4,c]
			X[3,c]}
		\caption{Glossary of terms and notation.} \label{table:2} \\
		\toprule
		Notation & Meaning & Type \\
		\midrule
		$S$ &	Set of all possible (hidden) states. & Finite set of cardinality $m>0$.\\\addlinespace[0.3cm]
		
		$s_\tau$ & (Hidden) state at time $\tau$. In computations, if $s_\tau$ evaluates to the i$^{th}$ possible state, then interpret it as the i$^{th}$ unit vector in $\mathbb R^m$. &
		Random variable over $S$. \\\addlinespace[0.3cm]
		
		$s_{1:t}$ & Sequence of hidden states $s_1,...,s_t$. & Random variables over $S^t$. \\\addlinespace[0.3cm]
		
		$O$ &	Set of all possible outcomes. &	Finite set of cardinality $n>0$.\\\addlinespace[0.3cm]
		
		$o_\tau$& Outcome at time $\tau$. In computations, if $o_\tau$ evaluates to the j$^{th}$ possible outcome, then interpret it as the j$^{th}$ unit vector in $\mathbb R^n$. &
		Random variable over $O$. \\\addlinespace[0.3cm]
		
		$o_{1:t}$ & Sequence of outcomes $o_1,...,o_t$ & Random variables over $O^t$.\\\addlinespace[0.3cm]
		
		$T$ & Number of timesteps in a trial of observation epochs under the generative model. & Positive integer. \\\addlinespace[0.3cm]
		
		$U$ & Set of all possible actions. & Finite set.\\\addlinespace[0.3cm]
		
		$\Pi$ & Set of all allowable policies; i.e., sequences of actions. & Finite subset of $U^T$. \\\addlinespace[0.3cm]
		
		$\pi$ & Policy. & Random variable over $\Pi$, or element of $\Pi$ depending on context. \\\addlinespace[0.3cm]
		
		$Q$ & Approximate posterior distribution. & Probability distribution over the latent variables of the generative model.\\\addlinespace[0.3cm]
		
		$F, F_\pi$ & Variational free energy and variational free energy conditioned upon a policy. & Functionals of $Q$.\\\addlinespace[0.3cm]
		
		$G$ & Expected free energy. & Function defined on $\Pi$. \\\addlinespace[0.3cm]
		
		$Cat$ & Categorical distribution; probability distribution over a finite set assigning strictly positive probabilities. & Probability distribution over a finite set of cardinality $k$ with parameter space $\{x \in \mathbb R^{k} | x_i >0, \sum_i x_i =1\}$\\\addlinespace[0.3cm]
		
		$Dir$ & Dirichlet distribution (conjugate prior of the categorical distribution); probability distribution over the parameter space of the categorical distribution, parameterised by a vector of positive reals. &  Probability distribution over $\{x \in \mathbb R^{k} | x_i >0, \sum_i x_i =1\}$, itself parameterised by an element of $(\mathbb R_{>0})^k$. \\\addlinespace[0.3cm]
		
		$X_{\bullet i}, X_{ki}$& i$^{th}$ column and (k,i)$^{th}$ element of matrix $X$. & Matrix indexing convention. \\\addlinespace[0.3cm]
		
		$\cdot, \otimes, \odot, ^\odot$& Respectively inner product, Kronecker product, element-wise product and element-wise power. Following existing active inference literature, we adopt the convention $X \cdot Y:= X^T Y$ for matrices. &	Operation on vectors and matrices. \\\addlinespace[0.3cm]
		
		$A$ & Likelihood matrix. The probability of the state-outcome pair $o_\tau, s_\tau$ is given by $o_\tau \cdot A s_\tau$. & Random variable over the subset of $M_{n\times m}(\mathbb R)$ with columns in $\{x \in \mathbb R^{k} | x_i >0, \sum_i x_i =1\}$. \\\addlinespace[0.3cm]
		
		$B$ & Matrix of transition probabilities from one state to the next state given action $\pi_{\tau-1}$. The probability of possible state $s_\tau$, given $s_{\tau-1}$ and action $\pi_{\tau-1}$ is $s_\tau\cdot B_{\pi_{\tau-1}}s_{\tau -1}$. & Matrix in $M_{m\times m}(\mathbb R)$ with columns in $\{x \in \mathbb R^{k} | x_i >0, \sum_i x_i =1\}$. \\\addlinespace[0.3cm]
		
		$a, \bold a$& Parameters of prior and approximate posterior beliefs about $A$. & Matrices in $M_{n\times m}(\mathbb R_{>0})$. \\\addlinespace[0.3cm]
		
		$a_0, \bold a_0$& Matrices of the same size as $a, \bold a$, with homogeneous columns; any of its i$^{th}$ column elements are denoted by $a_{i0}, \bold a_{i0}$ and defined by $a_{i0} = \sum_{j=1}^n a_{ji}, \bold a_{i0}= \sum_{j=1}^n \bold a_{ji}$. & Matrices in $M_{n\times m}(\mathbb R_{>0})$. \\\addlinespace[0.3cm]
		
		$\log, \Gamma, \psi$ & Natural logarithm, gamma function and digamma function. By convention these functions are taken component-wise on vectors and matrices. &	Functions.\\\addlinespace[0.3cm]
		
		$\mathbb E_{P(X)}[f(X)]$ & Expectation of random variable $f(X)$ under a probability density $P(X)$, taken component-wise if $f(X)$ is a matrix. $\mathbb E_{P(X)}[f(X)] := \int f(X) P(X)\: dX$  &	Real-valued operator on random variables. \\\addlinespace[0.3cm]
		
		$\bold A$ & $\bold A := \E_{Q(A)}[A] = \bold a \odot \bold a_0^{\odot( -1)}$  & Matrix in $M_{n\times m}(\mathbb R_{>0})$. \\\addlinespace[0.3cm]
		
		$\textbf{log} \bold A$ & $\textbf{log}\bold A := \E_{Q(A)}[\log A] = \psi(\bold a)-\psi(\bold a_0)$. Note that $\textbf{log} \bold A \neq \log \bold A$! & Matrix in $M_{n\times m}(\mathbb R)$. \\\\\addlinespace[0.3cm]
		
		$\sigma$ & Softmax function or normalised exponential. $\sigma(x)_k = \frac{e^{x_k}}{\sum_i e^{x_i}}$ & Function $\mathbb R^k \to \{x \in \mathbb R^{k} | x_i >0, \sum_i x_i =1\}$\\\addlinespace[0.3cm]
		
		$\text H[P]$& Shannon entropy of a probability distribution $P$. Explicitly, $\text H[P]=\E_{P(x)}[-\log P(x)]$ &	Functional over probability distributions. \\ \addlinespace[0.15cm]
		
		\bottomrule
	\end{longtabu}

	As mentioned above, a substantial body of work justifies describing certain neuronal representations with discrete state-space generative models (e.g., \cite{zhangDiscreteFixedResolutionRepresentations2008,luckCapacityVisualWorking1997,teeInformationBrainRepresented2018}). Furthermore, it has been long known that -- at the level of neuronal populations -- computations occur periodically (i.e., in distinct and sometimes nested oscillatory bands). Similarly, there is evidence for sequential computation in a number of processes (e.g., attention \cite{buschmanShiftingSpotlightAttention2010,duncanDirectMeasurementAttentional1994,landauAttentionSamplesStimuli2012}, visual perception \cite{hanslmayrPrestimulusOscillatoryPhase2013,rollsProcessingSpeedCerebral1994}) and at different levels of the neuronal hierarchy \cite{fristonDeepTemporalModels2018,fristonHierarchicalModelsBrain2008}, in line with ideas from hierarchical predictive processing \cite{chaoLargeScaleCorticalNetworks2018,iglesiasHierarchicalPredictionErrors2013}. This accommodates the fact that visual saccadic sampling of observations occurs at a frequency of approximately $4$Hz \cite{parrDiscreteContinuousBrain2018}. The relatively slow presentation of a discrete sequence of observations enables inferences to be performed in peristimulus time by (much) faster neuronal dynamics.
	
	Active inference, implicitly, accounts for fast and slow neuronal dynamics. At each time-step the agent observes an outcome, from which it infers the past, present and future (hidden) states through perception. This underwrites a plan into the future, by evaluating (the expected free energy of) possible policies. The inferred (best) policies specify the most likely action, which is executed. At a slower timescale, parameters encoding the contingencies of the world (e.g., $A$), are inferred. This is referred to as learning. Even more slowly, the structure of the generative model is updated to better account for available observations -- this is called structure learning. The following sections elucidate these aspects of the active inference process theory.
	
	This paper will be largely concerned with deriving and interpreting the inferential dynamics that agents might implement using the generative model in Figure \ref{fig:gen mod}. We leave the discussion of more complex models to Appendix \ref{appendix: more complex models}, since the derivations are analogous in those cases.
	
	\section{Variational Bayesian inference}
	\label{sec: VB}
	
	\subsection{Free energy and model evidence}
	
	Variational Bayesian inference rests upon minimisation of a quantity called (variational) free energy, which measures the improbability (i.e., the surprise) of sensory observations, under a generative model. Simultaneously, variational free energy minimisation is a statistical inference technique that enables the approximation of the posterior distribution in Bayes rule. In machine learning, this is known as variational Bayes \cite{bishopPatternRecognitionMachine2006,jordanIntroductionVariationalMethods1998,bealVariationalAlgorithmsApproximate2003,bleiVariationalInferenceReview2017}.
	Active inference agents minimise variational free energy, enabling concomitant maximisation of their model evidence and inference of the latent variables of their generative model. In the following, we consider a particular time point to be given $t\in\{1,...,T\}$, whence the agent has observed a sequence of outcomes $o_{1:t}$. The posterior about the latent causes of sensory data is given by Bayes rule:
	
	\begin{equation}
		\label{eq:bayes rule}
		P(s_{1:T}, A, \pi |o_{1:t})= \frac{P(o_{1:t} |s_{1:T}, A, \pi )P(s_{1:T}, A, \pi )}{P(o_{1:t})}
	\end{equation}

	Computing the posterior distribution requires computing the model evidence $P(o_{1:t})=\sum_{\pi \in \Pi }\sum_{s_{1:T}\in S^T}\int P(o_{1:t} ,s_{1:T}, A, \pi ) \: dA$, which is intractable for complex generative models embodied by biological and artificial systems \cite{fristonHierarchicalModelsBrain2008} -- a well-known problem in Bayesian statistics.
	An alternative to computing the exact posterior distribution is to optimise an approximate posterior distribution over latent causes $Q(s_{1:T}, A, \pi)$, by minimising the Kullback-Leibler (KL) divergence \cite{kullbackInformationSufficiency1951} ($D_{KL}$) -- a non-negative measure of discrepancy between probability distributions. We can use the definition of the KL divergence and Bayes rule to arrive at the variational free energy $F$, which is a functional of approximate posterior beliefs:
	
	\begin{equation}
		\label{eq: vfe}
		\begin{split}
			0 &\leq D_{KL}[Q(s_{1:T}, A, \pi)||P(s_{1:T}, A, \pi |o_{1:t})] \\
			&= \E_{Q(s_{1:T}, A, \pi)}[\log Q(s_{1:T}, A, \pi) - \log P(s_{1:T}, A, \pi |o_{1:t})] \\
			&= \E_{Q(s_{1:T}, A, \pi)}[\log Q(s_{1:T}, A, \pi) - \log P(o_{1:t},s_{1:T}, A, \pi) + \log P(o_{1:t})] \\
			&= \underbrace{\E_{Q(s_{1:T}, A, \pi)}[\log Q(s_{1:T}, A, \pi) - \log P(o_{1:t},s_{1:T}, A, \pi)]}_{F[Q(s_{1:T}, A, \pi)]} + \log P(o_{1:t}) \\
			\Rightarrow & -\log P(o_{1:t}) \leq F[Q(s_{1:T}, A, \pi)]
		\end{split}
	\end{equation}
	
	From \eqref{eq: vfe}, one can see that by varying $Q$ to minimise the variational free energy enables us to approximate the true posterior, while simultaneously ensuring that surprise remains low. This means that variational free energy minimising agents, simultaneously, infer the latent causes of their observations and maximise the evidence for their generative model.
	To aid intuition, the variational free energy can be rearranged into complexity and accuracy:
	
	\begin{equation}
		\label{eq: complexity accuracy}
		F[Q(s_{1:T}, A, \pi)] = \underbrace{D_{KL}[Q(s_{1:T}, A, \pi)||P(s_{1:T}, A, \pi)]}_{\text{Complexity}}-\underbrace{\E_{Q(s_{1:T}, A, \pi)}[\log P(o_{1:t}|s_{1:T}, A, \pi)]}_{\text{Accuracy}}
	\end{equation}
	
\begin{figure}
		\centering
		\includegraphics[width=\textwidth]{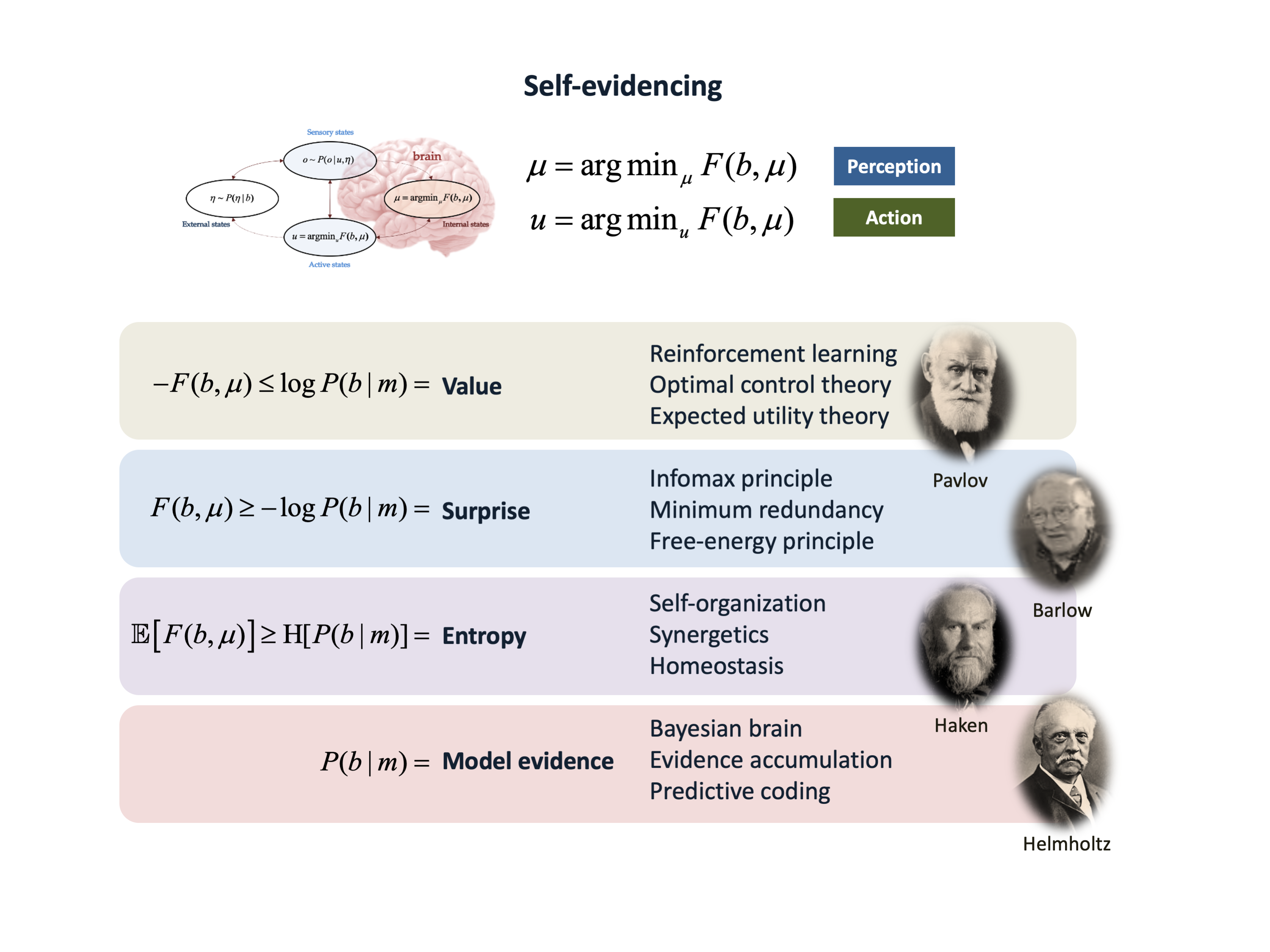}
		\caption{\textbf{Markov blankets and self-evidencing.}
			This schematic illustrates the various interpretations of minimising variational free energy. Recall that the existence of a Markov blanket implies a certain lack of influences among internal, blanket and external states. These independencies have an important consequence; internal and active states are the only states that are not influenced by external states, which means their dynamics (i.e., perception and action) are a function of, and only of, particular states (i.e., internal, sensory and active states); here, the variational (free energy) bound on surprise. This surprise has a number of interesting interpretations. Given it is the negative log probability of finding a particle or creature in a particular state, minimising surprise corresponds to maximising the value of a particle’s state. This interpretation is licensed by the fact that the states with a high probability are, by definition, attracting states. On this view, one can then spin-off an interpretation in terms of reinforcement learning \cite{bartoReinforcementLearningIntroduction1992}, optimal control theory \cite{todorovOptimalFeedbackControl2002} and, in economics, expected utility theory \cite{bossaertsBehaviouralEconomicsNeuroeconomics2015}. Indeed, any scheme predicated on the optimisation of some objective function can now be cast in terms of minimising surprise -- in terms of perception and action (i.e., the dynamics of internal and active states) -- by specifying these optimal values to be the agent's preferences. The minimisation of surprise (i.e., self-information) leads to a series of influential accounts of neuronal dynamics; including the principle of maximum mutual information \cite{opticanTemporalEncodingTwodimensional1987,linskerPerceptualNeuralOrganization1990}, the principles of minimum redundancy and maximum efficiency \cite{barlowPossiblePrinciplesUnderlying1961} and the free energy principle \cite{fristonFreeEnergyPrinciple2006}. Crucially, the average or expected surprise (over time or particular states of being) corresponds to entropy. This means that action and perception look as if they are minimising entropy. This leads us to theories of self-organisation, such as synergetics in physics \cite{hakenSynergeticsIntroductionNonequilibrium1978,kauffmanOriginsOrderSelforganization1993,nicolisSelforganizationNonequilibriumSystems1977} or homeostasis in physiology \cite{ashbyPrinciplesSelfOrganizingDynamic1947,bernardLecturesPhenomenaLife1974,conantEveryGoodRegulator1970}. Finally, the probability of any blanket states given a Markov blanket ($m$) is, on a statistical view, model evidence \cite{mackayInformationTheoryInference2003,mackayFreeEnergyMinimization1995}. This means that all the above formulations are internally consistent with things like the Bayesian brain hypothesis, evidence accumulation and predictive coding; most of which inherit from Helmholtz's motion of unconscious inference \cite{helmholtzHelmholtzTreatisePhysiological1962}, later unpacked in terms of perception as hypothesis testing in 20th century psychology \cite{gregoryPerceptionsHypotheses1980} and machine learning \cite{dayanHelmholtzMachine1995}.}
		\label{fig: VFE interpretations}
	\end{figure}

	The first term of \eqref{eq: complexity accuracy} can be regarded as complexity: a simple explanation for observable data $Q$, which makes few assumptions over and above the prior (i.e., with KL divergence close to zero), is a good explanation. In other words, a good explanation is an accurate account of some data that requires minimal movement for updating of prior to posterior beliefs (c.f., Occam's principle). The second term is accuracy; namely, the probability of the data given posterior beliefs about model parameters $Q$. In other words, how well the generative model fits the observed data. The idea that neural representations weigh complexity against accuracy underwrites the imperative to find the most accurate explanation for sensory observations that is minimally complex, which has been leveraged by things like Horace Barlow’s principle of minimum redundancy \cite{barlowRedundancyReductionRevisited2001} and subsequently supported empirically \cite{danEfficientCodingNatural1996,lewickiEfficientCodingNatural2002,olshausenSparseCodingSensory2004,olshausenNewWindowSound2002}. Figure \ref{fig: VFE interpretations} illustrates the various implications of minimising free energy.
	
	\subsection{On the family of approximate posteriors}
	
	The goal is now to minimise variational free energy with respect to $Q$. To obtain a tractable expression for the variational free energy, we need to assume a certain simplifying factorisation of the approximate posterior. There are many possible forms \cite{parrNeuronalMessagePassing2019,yedidiaConstructingFreeEnergyApproximations2005,heskesConvexityArgumentsEfficient2006} (e.g., mean-field, marginal, Bethe), each of which trades off the quality of the inferences with the complexity of the computations involved. For the purpose of this paper we use a particular (structured) mean-field approximation (c.f., Figure \ref{fig:gen mod}):
	
	\begin{equation}
		\label{eq: mean field approx}
		Q(s_{1:T},A,\pi) = Q(A)Q(\pi) \prod_{\tau =1}^T Q(s_\tau|\pi)
	\end{equation}
	
	This choice is driven by didactic purposes and since this factorisation has been used extensively in the active inference literature (e.g., \cite{fristonDeepTemporalModels2018,fristonGraphicalBrainBelief2017,fristonActiveInferenceProcess2017}). However, the most recent software implementation of active inference (i.e., \texttt{spm\_MDP\_VB\_X.m}) employs a marginal approximation \cite{parrNeuronalMessagePassing2019,parrComputationalNeurologyActive2019}, which retains the simplicity and biological interpretation of the neuronal dynamics afforded by the mean-field approximation, while approximating the more accurate inferences of the Bethe approximation. For these reasons, the marginal free energy currently stands as the most biologically plausible.
	
	\subsection{Computing the variational free energy}
	
	The next sections focus on producing biologically plausible neuronal dynamics that perform perception and learning based on variational free energy minimisation. To enable this, we first compute variational the free energy, using the factorisations of the generative model and approximate posterior (c.f., Figure \ref{fig:gen mod}):
	
	\begin{equation}
		\label{eq: computing VFE}
		\begin{split}
			F[Q(s_{1:T},A, \pi)]&= \E_{Q(s_{1:T},A, \pi)}[\log Q(s_{1:T},A, \pi) - \log P(o_{1:t},s_{1:T},A, \pi)] \\
			&= \E_{Q(s_{1:T},A, \pi)}[\log Q(A) +  \log Q(\pi)+\sum_{\tau = 1}^T  \log Q(s_\tau|\pi) \\
			&-\log P(A) -\log P(\pi)- \log P(s_1)- \sum_{\tau =2}^T\log P(s_\tau |s_{\tau-1},\pi)-\sum_{\tau = 1}^t \log P(o_{\tau}|s_{\tau},A)] \\
			&=D_{KL}[Q(A)||P(A)]+D_{KL}[Q(\pi)||P(\pi)]+\E_{Q(\pi)}[F_\pi[Q(s_{1:T}|\pi)]]
		\end{split}
	\end{equation}
	where
	\begin{equation}
		\label{eq: VFE_pi}
		\begin{split}
			F_\pi[Q(s_{1:T}|\pi)] & := \sum_{\tau =1}^T\E_{Q(s_\tau |\pi)}[\log Q(s_\tau |\pi)] -\sum_{\tau = 1}^t \E_{Q(s_\tau |\pi)Q(A)}[\log P(o_{\tau}|s_{\tau},A)] \\
			&-\E_{Q(s_1 |\pi)}[\log P(s_1)]-\sum_{\tau =2}^T \E_{Q(s_\tau |\pi)Q(s_{\tau-1} |\pi)}[\log P(s_\tau |s_{\tau-1},\pi)]
		\end{split}
	\end{equation}
	
	is the variational free energy conditioned upon a policy. This is the same quantity that we would have obtained by omitting $A$ and conditioning all probability distributions in the numerators of \eqref{eq:bayes rule} by $\pi$. In the next section, we will see how perception can be framed in terms of variational free energy minimisation.
	
	\section{Perception}
	
	In active inference, perception is equated with state estimation \cite{fristonActiveInferenceProcess2017} (e.g., inferring the temperature from the sensation of warmth), consistent with the idea that perceptions are hypotheses \cite{gregoryPerceptionsHypotheses1980}. To infer the (past, present and future) states of the environment, an agent must minimise the variational free energy with respect to $Q(s_{1:T}|\pi)$ for each policy $\pi$. This provides the agent’s inference over hidden states, contingent upon pursuing a given policy. Since the only part of the free energy that depends on $Q(s_{1:T}|\pi)$ is $F_\pi$, the agent must simply minimise $F_\pi$. 
	Substituting $Q(s_{\tau}|\pi)$ by their sufficient statistics (i.e., parameters $\bold s_{\pi \tau}$), $F_\pi$ becomes a function of those parameters. This enables us to rewrite \eqref{eq: VFE_pi}, conveniently in matrix form:
	
	\begin{equation}
		\begin{split}
			F_\pi(\bold s_{\pi 1},...,\bold s_{\pi T}) &= \sum_{\tau = 1}^T \bold s_{\pi \tau } \cdot \log \bold s_{\pi \tau } 
			- \sum_{\tau = 1}^t o_\tau \cdot \textbf{log} \bold A \bold s_{\pi \tau }\\
			&- \bold s_{\pi 1 } \log D- \sum_{\tau = 2}^T \bold s_{\pi \tau } \cdot \log (B_{\pi_{\tau-1}})\bold s_{\pi \tau-1 }
		\end{split}
	\end{equation}
	
	This enables to compute the variational free energy gradients \cite{petersenMatrixCookbook}:
	\begin{equation}
		\label{eq:vfe gradients perception}
		\begin{split}
			\nabla_{\bold s_{\pi\tau}} F_\pi(\bold s_{\pi 1},...,\bold s_{\pi T}) = \vec {1} +\log  s_{\pi\tau} -
			\begin{cases}
				o_\tau \cdot \textbf{log} \bold A+ \bold s_{\pi\tau+1} \cdot \log (B_{\pi_{\tau}}) + \log D \quad  \text{if } \tau = 1 \\
				o_\tau \cdot \textbf{log} \bold A+ \bold s_{\pi\tau+1} \cdot \log (B_{\pi_{\tau}}) +\log (B_{\pi_{\tau-1}}) \bold s_{\pi\tau-1} \quad \text{if } 1<\tau \leq t  \\
				\bold s_{\pi\tau+1} \cdot \log (B_{\pi_{\tau}}) +\log (B_{\pi_{\tau-1}}) \bold s_{\pi\tau-1} \quad \text{if }  \tau > t
			\end{cases}
		\end{split}
	\end{equation}
	
	The neuronal dynamics are given by a gradient descent on free energy \cite{fristonActiveInferenceProcess2017}, with state-estimation expressed as a softmax function of accumulated (negative) free energy gradients. The constant term $\vec{1}$ is generally omitted since the softmax function removes it anyway. This enables us to equate the gradient with a prediction error.
	
	\begin{equation}
		\label{eq: perceptual dynamics}
		\begin{split}
			\dot v (\bold s_{\pi 1},...,\bold s_{\pi T})&=-\nabla_{\bold s_{\pi\tau}} F_\pi(\bold s_{\pi 1},...,\bold s_{\pi T}) \\
			\bold s_{\pi\tau} = \sigma (v)
		\end{split}
	\end{equation}
	
	The softmax function – a generalisation of the sigmoid to vector inputs – is a natural choice as the variational free energy gradient is a logarithm and the components of $\bold s_{\pi\tau}$ must sum to one.
	
	\subsection{Plausibility of neuronal dynamics}
	
	The temporal dynamics expressed in \eqref{eq: perceptual dynamics} unfold at a much faster timescale than the sampling of new observations (i.e., within timesteps) and correspond to fast neuronal processing in peristimulus time. This is consistent with behaviour-relevant computations at frequencies that are higher than the rate of visual sampling (e.g., working memory \cite{lundqvistGammaBetaBursts2016}, visual stimulus perception in humans \cite{hanslmayrPrestimulusOscillatoryPhase2013} and macaques \cite{rollsProcessingSpeedCerebral1994}).
	
	Furthermore, these dynamics are consistent with predictive processing \cite{raoPredictiveCodingVisual1999,bastosCanonicalMicrocircuitsPredictive2012} -- since active inference prescribes dynamics that minimise prediction error -- although they generalise it to a wide range of generative models. Note that, while also a variational free energy gradient, this sort of prediction error is not the same as that given by predictive coding schemes (which rely upon a different sort of generative model) \cite{buckleyFreeEnergyPrinciple2017,fristonVariationalFreeEnergy2007,bogaczTutorialFreeenergyFramework2017}.
	
	Just as neuronal dynamics involve translation from post-synaptic potentials to firing rates, \eqref{eq: perceptual dynamics} involves translating from a vector of real numbers ($v$), to a vector whose elements are bounded between zero and one ($\bold s_{\pi\tau}$); via the softmax function. As a result, it is natural to interpret the components of $v$ as the average membrane potential of distinct neural populations, and $\bold s_{\pi\tau}$ as the average firing rate of those populations, which is bounded thanks to neuronal refractory periods. This is consistent with mean-field formulations of neural population dynamics, in that the average firing rate of a neuronal population follows a sigmoid function of the average membrane potential \cite{marreirosPopulationDynamicsVariance2008,decoDynamicBrainSpiking2008,moranNeuralMassesFields2013}.
	Using the fact that a softmax function is a generalisation of the sigmoid to vector inputs -- here the average membrane potentials of coupled neuronal populations -- it follows that their average firing follows a softmax function of their average potential. In this context, the softmax function may be interpreted as performing lateral inhibition, which can be thought of as leading to narrower tuning curves of individual neurons and thereby sharper inferences \cite{vonbekesySensoryInhibition1967}. Importantly, this tells us that state-estimation can be performed in parallel by different neuronal populations, and a simple neuronal architecture is sufficient to implement these dynamics (see Figure 6 in \cite{parrNeuronalMessagePassing2019}).
	
	Lastly, interpreting the dynamics in this way has a degree of face validity, as it enables us to synthesise a wide-range of biologically plausible electrophysiological responses; including repetition suppression, mismatch negativity, violation responses, place-cell activity, phase precession, theta sequences, theta-gamma coupling, evidence accumulation, race-to-bound dynamics and transfer of dopamine responses \cite{schwartenbeckDopaminergicMidbrainEncodes2015,fristonActiveInferenceProcess2017}.
	
	The neuronal dynamics for state estimation coincide with variational message passing \cite{winnVariationalMessagePassing2005,dauwelsVariationalMessagePassing2007}: a widely used algorithm for approximate Bayesian inference. This is an important result, since it shows that variational message passing emerges under active inference using a particular mean-field approximation. If one were to use the Bethe approximation, the corresponding dynamics coincide with belief propagation \cite{bishopPatternRecognitionMachine2006,loeligerIntroductionFactorGraphs2004,parrNeuronalMessagePassing2019,schwobelActiveInferenceBelief2018,yedidiaConstructingFreeEnergyApproximations2005}, another widely used algorithm for approximate inference. This offers a formal connection between active inference and message passing interpretations of neuronal dynamics \cite{fristonGraphicalBrainBelief2017,dauwelsMeasuringNeuralSynchrony2007,georgeBeliefPropagationWiring2005}. In the next section, we examine planning, decision-making and action selection.

	\section{Planning, decision-making and action selection}
	
	So far, we have focused on optimising beliefs about hidden states under a particular policy by minimising a variational free energy functional of an approximate posterior over hidden states, under each policy.
	
	In this section, we explain how planning and decision-making arise as a minimisation of expected free energy -- a function scoring the goodness of each possible future course of action. We briefly motivate how the expected free energy arises from first-principles. This allows us to frame decision-making and action-selection in terms of expected free energy minimisation. Finally, we conclude by discussing the computational cost of planning into the future.
	
	\subsection{Planning and decision-making}
	
	At the heart of active inference, is a description of agents that strive to attain a target distribution specifying the range of preferred states of being, given a sufficient amount of time. To work towards reaching these preferences, agents select policies $Q(\pi)$, such that their predicted states $Q(s_\tau,A)$ at some future time point $\tau > t$ (usually, the time horizon of a policy $T$) reach the preferred states $P(s_\tau,A)$, which are specified by the generative model. These considerations allow us to show in Appendix \ref{Appendix:steady-state lemma} that the requisite approximate posterior over policies $Q(\pi)$ is a softmax function of the negative \textit{expected free energy} $G$\footnote{A more complete treatment may include priors over policies – usually denoted by $E$ – and the evidence for a policy afforded by observed outcomes (usually denoted by $\bold F$). These additional terms supplement the expected free energy, leading to an approximate posterior of the form $\sigma (- \log E - \bold F - \bold G)$ \cite{fristonDeepTemporalModels2018}.}:
	
	\begin{equation}
		\label{eq: approx post policies}
		\begin{split}
			Q(\pi) &=\sigma(-G(\pi)) \\
			G(\pi) &= \underbrace{ D_{KL}[Q(s_\tau, A|\pi)||P(s_\tau, A)]}_{Risk}- \underbrace{\mathbb E_{Q(s_\tau, A|\pi)P(o_\tau|s_\tau, A)}[ \log P(o_\tau |s_\tau, A)]}_{Ambiguity}
		\end{split}
	\end{equation}
	
	This means that the most likely (i.e., best) policies minimise expected free energy. This ensures that future courses of action are exploitative (i.e., \textit{risk} minimising) and explorative (i.e., \textit{ambiguity} minimising). In particular, the expected free energy specifies the optimal balance between goal-seeking and itinerant novelty-seeking behaviour, given some prior preferences or goals. Note that the ambiguity term rests on an expectation over fictive (i.e., predicted) outcomes under beliefs about future states. This means that optimising beliefs about future states during perception is crucial to accurately predict future outcomes during planning. In summary, planning and decision-making respectively correspond to evaluating the expected free energy of different policies, which scores their goodness in relation to prior preferences and forming approximate posterior beliefs about policies.
	
	\subsection{Action selection, policy-independent state-estimation}
	
	Approximate posterior beliefs about policies allows to obtain the most plausible action as the most likely under all policies -- this can be expressed as a Bayesian model average
	
	\begin{equation}
		\label{eq: action selection}
		u_t = \arg \max_{u \in U } \left (\sum_{\pi \in \Pi} \delta_{u, \pi_t} Q(\pi)\right)
	\end{equation}
	
	where $\delta$ is the Kronecker delta. In addition, we obtain a policy independent state-estimation at any time point $\tau \in \{1,...,T\}$, as a Bayesian model average of approximate posterior beliefs about hidden states under policies:
	
	\begin{equation}
		\label{eq: policy indep}
		\begin{split}
			Q(s_\tau) &= \sum_{\pi \in \Pi} Q(s_\tau |\pi)Q(\pi)\\
			\iff \bold s_\tau &= \sum_{\pi \in \Pi} \bold s_{\pi\tau} Q(\pi)
		\end{split}
	\end{equation}
	
	Note that these Bayesian model averages may be implemented by neuromodulatory mechanisms \cite{fitzgeraldModelAveragingOptimal2014}.
	
	\subsection{Biological plausibility}
	
	Winner take-all architectures of decision-making are already commonplace in computational neuroscience (e.g., models of selective attention and recognition \cite{carpenterMassivelyParallelArchitecture1987,ittiModelSaliencybasedVisual1998}, hierarchical models of vision \cite{riesenhuberHierarchicalModelsObject1999}).
	This is nice, since the softmax function in \eqref{eq: approx post policies} can be seen as providing a biologically plausible \cite{marreirosPopulationDynamicsVariance2008,decoDynamicBrainSpiking2008,moranNeuralMassesFields2013}, smooth approximation to the maximum operation, which is known as soft winner take-all \cite{maassComputationalPowerWinnerTakeAll2000}. In fact, the generative model, presented in Figure \ref{fig:gen mod}, can be naturally extended such that the approximate posterior contains an (inverse) temperature parameter $\gamma$ multiplying the expected free energy inside the softmax function (see Appendix \ref{appendix: gamma}). This temperature parameter regulates how precisely the softmax approximates the maximum function, thus recovering winner take-all architectures for high parameter values (technically, this converts Bayesian model averaging into Bayesian model selection, where the policy corresponds to a model of what the agent is doing). This parameter, regulating precision of policy selection, has a clear biological interpretation in terms of confidence encoded in dopaminergic firing \cite{fitzgeraldDopamineRewardLearning2015,fristonAnatomyChoiceDopamine2014,schwartenbeckDopaminergicMidbrainEncodes2015,fristonActiveInferenceProcess2017}. Interestingly, Daw and colleagues \cite{dawCorticalSubstratesExploratory2006} uncovered evidence in favour of a similar model employing a softmax function and temperature parameter in human decision-making.
	
	\subsection{Pruning of policy trees}
	
	From a computational perspective, planning (i.e., computing the expected free energy) for each possible policy can be cost-prohibitive, due do the combinatorial explosion in the number of sequences of actions when looking deep into the future. There has been work in understanding how the brain finesses this problem \cite{huysBonsaiTreesYour2012}, which suggests a simple answer: during mental planning, humans stop evaluating a policy as soon as they encounter a large loss (i.e., a high value of the expected free energy that renders the policy highly implausible). In active inference this corresponds to using an Occam window; that is, we stop evaluating the expected free energy of a policy if it becomes much higher than the best (smallest expected free energy) policy -- and set its approximate posterior probability to an arbitrarily low value accordingly. This biologically plausible pruning strategy drastically reduces the number of policies one has to evaluate exhaustively.
	
	Although effective and biologically plausible, the Occam window for pruning policy trees cannot deal with large policy spaces that ensue with deep policy trees and long temporal horizons. This means that pruning can only partially explain how biological organisms perform deep policy searches. Further research is needed to characterise the processes in which biological agents reduce large policy spaces to tractable subspaces. One explanation -- for the remarkable capacity of biological agents to evaluate deep policy trees -- rests on deep (hierarchical) generative models, in which policies operate at each level. These deep models enable long-term policies, modelling slow transitions among hidden states at higher levels in the hierarchy, to contextualise faster state transitions at subordinate levels (see Appendix A). The resulting (semi Markovian) process can then be specified in terms of a hierarchy of limited horizon policies that are nested over temporal scales; c.f., motor chunking \cite{dehaeneNeuralRepresentationSequences2015,fonollosaLearningChunkingSequences2015,harunoHierarchicalMOSAICMovement2003}.
	
	\subsection{Discussion of the action-perception cycle}
	
	Minimising variational and expected free energy are complementary and mutually beneficial processes. Minimisation of variational free energy ensures that the generative model is a good predictor of its environment; this allows the agent to accurately plan into the future by evaluating expected free energy, which in turn enables it to realise its preferences. In other words, minimisation of variational free energy is a vehicle for effective planning and reaching preferences via the expected free energy; in turn, reaching preferences minimises the expected surprise of future states of being.
	
	In conclusion, we have seen how agents plan into the future and make decisions about the best possible course of action. This concludes our discussion of the action-perception cycle. In the next section, we examine expected free energy in greater detail. Then, we will see how active agents can learn the contingencies of the environment and the structure of their generative model at slower timescales.

	\section{Properties of the expected free energy}
	\label{sec: efe}
	
	The expected free energy is a fundamental construct of interest. In this section, we unpack its main features and highlight its importance in relation to many existing theories in neurosciences and engineering.
	
	The expected free energy of a policy can be unpacked in a number of ways. Perhaps the most intuitive is in terms of risk and ambiguity:
	
	\begin{equation}
		G(\pi) = \underbrace{D_{KL}[Q(s_\tau, A|\pi)||P(s_\tau, A)]}_{\text{Risk}}+ \underbrace{\mathbb E_{Q(s_\tau, A|\pi)}[\text{H} [ P(o_\tau |s_\tau, A)]]}_{\text{Ambiguity}}
	\end{equation}
	
	This means that policy selection minimises risk and ambiguity. Risk, in this setting, is simply the difference between predicted and prior beliefs about final states. In other words, policies will be deemed more likely if they bring about states that conform to prior preferences. In the optimal control literature, this part of expected free energy underwrites KL control \cite{todorovGeneralDualityOptimal2008,vandenbroekRiskSensitivePath2010}. In economics, it leads to risk sensitive policies \cite{flemingRisksensitiveControlOptimal2002}. Ambiguity reflects the uncertainty about future outcomes, given hidden states. Minimising ambiguity therefore corresponds to choosing future states that generate unambiguous and informative outcomes (e.g., switching on a light in the dark).
	
	We can express the expected free energy of a policy as a bound on information gain and expected log (model) evidence (a.k.a., Bayesian risk): 
	
	\begin{equation}
		\label{eq: efe inf gain exp log evidence}
		\begin{split}
			G(\pi)
			&= \underbrace{\mathbb E_Q[D_{KL}[Q(s_\tau, A|o_\tau, \pi)||P(s_\tau, A|o_\tau)]]}_{\text{Expected evidence bound}}- \underbrace{\mathbb E_Q[\log P(o_\tau)]}_{\text{Expected log evidence}} \\
			&- \underbrace{\mathbb E_Q [D_{KL}[Q(s_\tau,A |o_\tau, \pi)||Q(s_\tau ,A|\pi)]]}_{\text{Expected information gain}} \\
			&\geq - \underbrace{\mathbb E_Q[\log P(o_\tau)]}_{\text{Expected log evidence}}- \underbrace{\mathbb E_Q [D_{KL}[Q(s_\tau,A |o_\tau, \pi)||Q(s_\tau,A |\pi)]]}_{\text{Expected information gain}} 
		\end{split}
	\end{equation}
	
	The first term in \eqref{eq: efe inf gain exp log evidence} is the expectation of log evidence under beliefs about future outcomes, while the second ensures that this expectation is maximally informed, when outcomes are encountered. Collectively, these two terms underwrite the resolution of uncertainty about hidden states (i.e., information gain) and outcomes (i.e., expected surprise) in relation to prior beliefs.
	
	When the agent's preferences are expressed in terms of outcomes (c.f., Figure \ref{fig:gen mod}), it is useful to express risk in terms of outcomes, as opposed to hidden states. This is most useful when the generative model is not known or during structure learning, when the state-space evolves over time. In these cases, the risk over hidden states can be replaced risk over outcomes by assuming the KL divergence between the predicted and true posterior (under expected outcomes) is small:
	
	\begin{equation}
		\label{eq:pref outcomes}
		\begin{split}
			\underbrace{D_{KL}[Q(s_\tau, A|\pi)||P(s_\tau, A)]}_{\text{Risk (states)}} &=\underbrace{D_{KL}[Q(o_\tau|\pi)||P(o_\tau)]}_{\text{Risk (outcomes)}}+\underbrace{\mathbb E_{Q(o_\tau |\pi)}[D_{KL}[Q(s_\tau, A|o_\tau, \pi)||P(s_\tau, A|o_\tau)]]}_{\approx 0} \\
			&\approx \underbrace{D_{KL}[Q(o_\tau|\pi)||P(o_\tau)]}_{\text{Risk (outcomes)}}
		\end{split}
	\end{equation}
	
	This divergence constitutes an expected evidence bound that also appears if we express expected free energy in terms of intrinsic and extrinsic value:
	
	\begin{equation}
		\begin{split}
			G(\pi) &= -\underbrace{\mathbb E_{Q(o_\tau |\pi)}[\log P(o_\tau)]}_{\text{Extrinsic value}}+ \underbrace{\mathbb E_{Q(o_\tau|\pi)}[D_{KL}[Q(s_\tau, A|o_\tau, \pi)||P(s_\tau, A|o_\tau)]]}_{\text{Expected evidence bound}} \\
			&-\underbrace{\mathbb E_{Q(o_\tau|\pi)}[D_{KL}[Q(s_\tau|o_\tau, \pi)||Q(s_\tau |\pi)]]}_{\text{Intrinsic value (states) or salience}}
			- \underbrace{\mathbb E_{Q(o_\tau,s_\tau|\pi)}[D_{KL}[Q(A|o_\tau, s_\tau, \pi)||Q(A)]]}_{\text{Intrinsic value (parameters) or novelty}}
		\end{split}
	\end{equation}
	
	Extrinsic value is just the expected value of log evidence, which can be associated with reward and utility in behavioural psychology and economics, respectively \cite{bartoNoveltySurprise2013,kauderGenesisMarginalUtility1953,schmidhuberFormalTheoryCreativity2010}. In this setting, extrinsic value is the negative of Bayesian risk \cite{bergerStatisticalDecisionTheory1985}, when reward is log evidence. The intrinsic value of a policy is its epistemic value or affordance \cite{fristonActiveInferenceEpistemic2015}. This is just the expected information gain afforded by a particular policy, which can be about hidden states (i.e., salience) or model parameters (i.e., novelty). It is this term that underwrites artificial curiosity \cite{schmidhuberDevelopmentalRoboticsOptimal2006}.
	
\begin{figure}
		\centering
		\includegraphics[width=\textwidth]{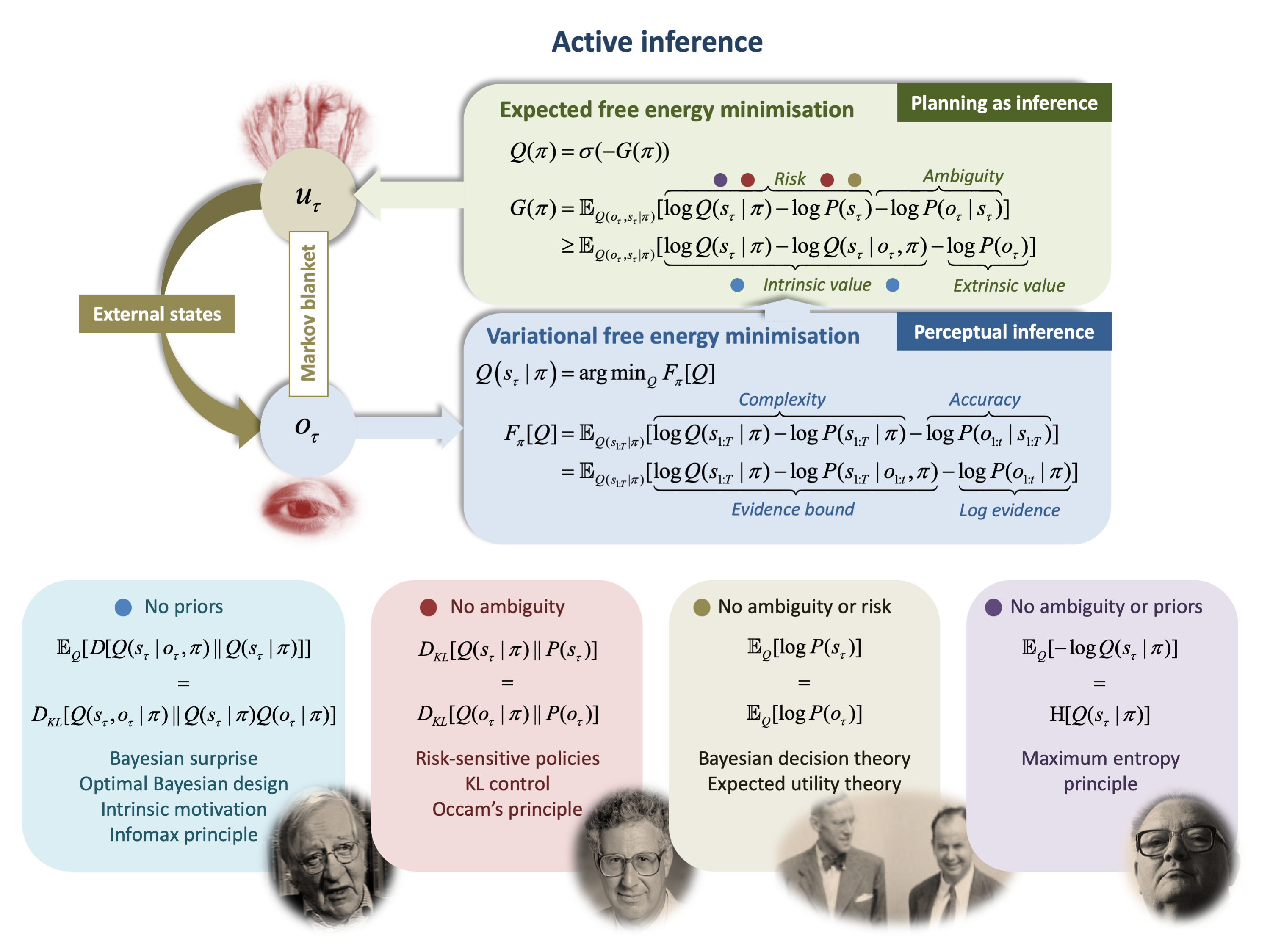}
		\caption{\textbf{Expected free energy.} This figure illustrates the various ways in which minimising expected free energy can be unpacked (omitting model parameters for clarity). The upper panel casts action and perception as the minimisation of variational and expected free energy, respectively. Crucially, active inference introduces beliefs over policies that enable a formal description of planning as inference \cite{kaplanPlanningNavigationActive2018,attiasPlanningProbabilisticInference2003,botvinickPlanningInference2012}. In brief, posterior beliefs about hidden states of the world, under plausible policies, are optimised by minimising a variational (free energy) bound on log evidence. These beliefs are then used to evaluate the expected free energy of allowable policies, from which actions can be selected \cite{fristonActiveInferenceProcess2017}. Crucially, expected free energy subsumes several special cases that predominate in the psychological, machine learning and economics literature. These special cases are disclosed when one removes particular sources of uncertainty from the implicit optimisation problem. For example, if we ignore prior preferences, then the expected free energy reduces to information gain \cite{mackayInformationTheoryInference2003,lindleyMeasureInformationProvided1956} or intrinsic motivation \cite{bartoNoveltySurprise2013,oudeyerWhatIntrinsicMotivation2009,deciIntrinsicMotivationSelfDetermination1985}. This is mathematically the same as expected Bayesian surprise and mutual information that underwrite salience in visual search \cite{ittiBayesianSurpriseAttracts2009,sunPlanningBeSurprised2011} and the organisation of our visual apparatus \cite{linskerPerceptualNeuralOrganization1990,opticanTemporalEncodingTwodimensional1987,barlowPossiblePrinciplesUnderlying1961,barlowInductiveInferenceCoding1974}. If we now remove risk but reinstate prior preferences, one can effectively treat hidden and observed (sensory) states as isomorphic. This leads to risk sensitive policies in economics \cite{kahnemanProspectTheoryAnalysis1988,flemingRisksensitiveControlOptimal2002} or KL control in engineering \cite{vandenbroekRiskSensitivePath2010}. Here, minimising risk corresponds to aligning predicted outcomes to preferred outcomes. If we then remove ambiguity and relative risk of action (i.e., intrinsic value), we are left with extrinsic value or expected utility in economics \cite{vonneumannTheoryGamesEconomic1944} that underwrites reinforcement learning and behavioural psychology \cite{bartoReinforcementLearningIntroduction1992}. Bayesian formulations of maximising expected utility under uncertainty is also known as Bayesian decision theory \cite{bergerStatisticalDecisionTheory1985}. Finally, if we just consider a completely unambiguous world with uninformative priors, expected free energy reduces to the negative entropy of posterior beliefs about the causes of data; in accord with the maximum entropy principle \cite{jaynesInformationTheoryStatistical1957}. The expressions for variational and expected free energy correspond to those described in the main text (omitting model parameters for clarity). They are arranged to illustrate the relationship between complexity and accuracy, which become risk and ambiguity, when considering the consequences of action. This means that risk-sensitive policy selection minimises expected complexity or computational cost. The coloured dots above the terms in the equations correspond to the terms that constitute the special cases in the lower panels.}
		\label{fig: EFE interpretations}
	\end{figure}

	Intrinsic value is also known as intrinsic motivation in neurorobotics \cite{bartoNoveltySurprise2013,oudeyerWhatIntrinsicMotivation2009,deciIntrinsicMotivationSelfDetermination1985}, the value of information in economics \cite{howardInformationValueTheory1966}, salience in the visual neurosciences and (rather confusingly) Bayesian surprise in the visual search literature \cite{ittiBayesianSurpriseAttracts2009,schwartenbeckExplorationNoveltySurprise2013,sunPlanningBeSurprised2011}. In terms of information theory, intrinsic value is mathematically equivalent to the expected mutual information between hidden states in the future and their consequences – consistent with the principles of minimum redundancy or maximum efficiency \cite{barlowPossiblePrinciplesUnderlying1961,barlowInductiveInferenceCoding1974,linskerPerceptualNeuralOrganization1990}. Finally, from a statistical perspective, maximising intrinsic value (i.e., salience and novelty) corresponds to optimal Bayesian design \cite{lindleyMeasureInformationProvided1956} and machine learning derivatives, such as active learning \cite{mackayInformationBasedObjectiveFunctions1992}. On this view, active learning is driven by novelty; namely, the information gain afforded model parameters, given future states and their outcomes. Heuristically, this curiosity resolves uncertainty about “what would happen if I did that” \cite{schmidhuberFormalTheoryCreativity2010}. Figure \ref{fig: EFE interpretations} illustrates the compass of expected free energy, in terms of its special cases; ranging from optimal Bayesian design through to Bayesian decision theory.
	
	\section{Learning}
	\label{sec: learning}
	
	In active inference, learning concerns the dynamics of synaptic plasticity, which are thought to encode beliefs about the contingencies of the environment \cite{fristonActiveInferenceProcess2017} (e.g., beliefs about $B$, in some settings, are thought to be encoded in recurrent excitatory connections in the prefrontal cortex \cite{parrPrefrontalComputationActive2019}). The fact that beliefs about matrices (e.g., $A$, $B$) may be encoded in synaptic weights conforms to connectionist models of brain function, as it offers a convenient way to compute probabilities, in the sense that the synaptic weights could be interpreted as performing matrix multiplication as in artificial neural networks, to predict; e.g., outcomes from beliefs about states, using the likelihood matrix $A$.
	
	These synaptic dynamics (e.g., long-term potentiation and depression) evolve at a slower timescale than action and perception, which is consistent with the fact that such inferences need evidence accumulation over multiple state-outcome pairs. For simplicity, we will assume the only variable that is learned is $A$, but what follows generalises to more complex generative models (c.f., Appendix \ref{appendix:learning B D}. Learning $A$ means that approximate posterior beliefs about $A$ follow a gradient descent on variational free energy. Seeing the variational free energy \eqref{eq: computing VFE} as a function of $\bold a$ (the sufficient statistic of $Q(A)$) we can write:
	
	\begin{equation}
		\label{eq: vfe learning}
		\begin{split}
			F(\bold a)&= D_{KL}[Q(A)||P(A)]-\sum_{\tau=1}^t \E_{Q(\pi)Q(s_\tau|\pi)Q(A)}[o_\tau\cdot \log (A) s_\tau] +\cdots \\
			&= D_{KL}[Q(A)||P(A)]-\sum_{\tau=1}^t o_\tau \cdot \textbf{log} \bold A \bold s_\tau+\cdots
		\end{split}
	\end{equation}
	
	Here, we ignore the terms in \eqref{eq: computing VFE} that do not depend on $Q(A)$, as these will vanish when we take the gradient. The KL-divergence between Dirichlet distributions is \cite{kurtKullbackLeiblerDivergenceTwo2013,pennyKLdivergenceNormalGamma2001}:	 
	
	\begin{equation}
		\label{eq: DKL A}
		\begin{split}
			D_{KL}[Q(A)||P(A)]&=\sum_{i=1}^m D_{KL}[Q(A_{\cdot i})||P(A_{\cdot i})] \\
			&= \sum_{i=1}^m \left( \log \Gamma(\bold a_{0i})- \sum_{k=1}^n \log \Gamma(\bold a_{ki}) -\log \Gamma( a_{0i})+ \sum_{k=1}^n \log \Gamma( a_{ki}) + (\bold a_{\bullet i}- a_{\bullet i})\cdot (\textbf{log}\bold A)_{\bullet i}\right) \\
			&= \sum_{i=1}^m \left( \log \Gamma(\bold a_{0i})- \sum_{k=1}^n \log \Gamma(\bold a_{ki}) -\log \Gamma( a_{0i})+ \sum_{k=1}^n \log \Gamma( a_{ki})\right) + (\bold a- a)\cdot \textbf{log}\bold A
		\end{split}
	\end{equation}
	
	Incorporating \eqref{eq: DKL A} in \eqref{eq: vfe learning}, we can take the gradient of the variational free energy with respect to $\textbf{log} \bold A$:
	
	\begin{equation}
		\label{eq: gradient vfe learning}
		\nabla_{\textbf{log}\bold A} F(\bold a)= \bold a -a -\sum_{\tau =1}^t o_\tau \otimes \bold s_\tau
	\end{equation}
	
	where $\otimes$ is the Kronecker (i.e., outer) product. This means that the dynamics of synaptic plasticity follow a descent on \eqref{eq: gradient vfe learning}:
	
	\begin{equation}
		\begin{split}
			\dot \rho(\bold a) &= -\nabla_{\textbf{log}\bold A} F(\bold a)\\
			&= -\bold a +a +\sum_{\tau =1}^t o_\tau \otimes \bold s_\tau
		\end{split}
	\end{equation}
	
	In computational terms, these are the dynamics for evidence accumulation of Dirichlet parameters at time $t$. Since synaptic plasticity dynamics occur at a much slower pace than perceptual inference, it is computationally much cheaper -- in numerical simulations -- to do a one-step belief update at the end of each trial of observation epochs. Explicitly, setting the free energy gradient to zero at the end of the trial gives the following update for Dirichlet parameters:
	
	\begin{equation}
		\label{eq: learning dirichlet one step}
		\bold a=  a +\sum_{\tau =1}^T o_\tau \otimes \bold s_\tau
	\end{equation}
	
	After which, the prior beliefs $P(A)$ are updated to the approximate posterior beliefs $Q(A)$ for the subsequent trial. Note that in particular, the update counts the number of times a specific mapping between states and observations has been observed. Interestingly, this is formally identical to associative or Hebbian plasticity.

\begin{figure}
		\centering
		\includegraphics[width=\textwidth]{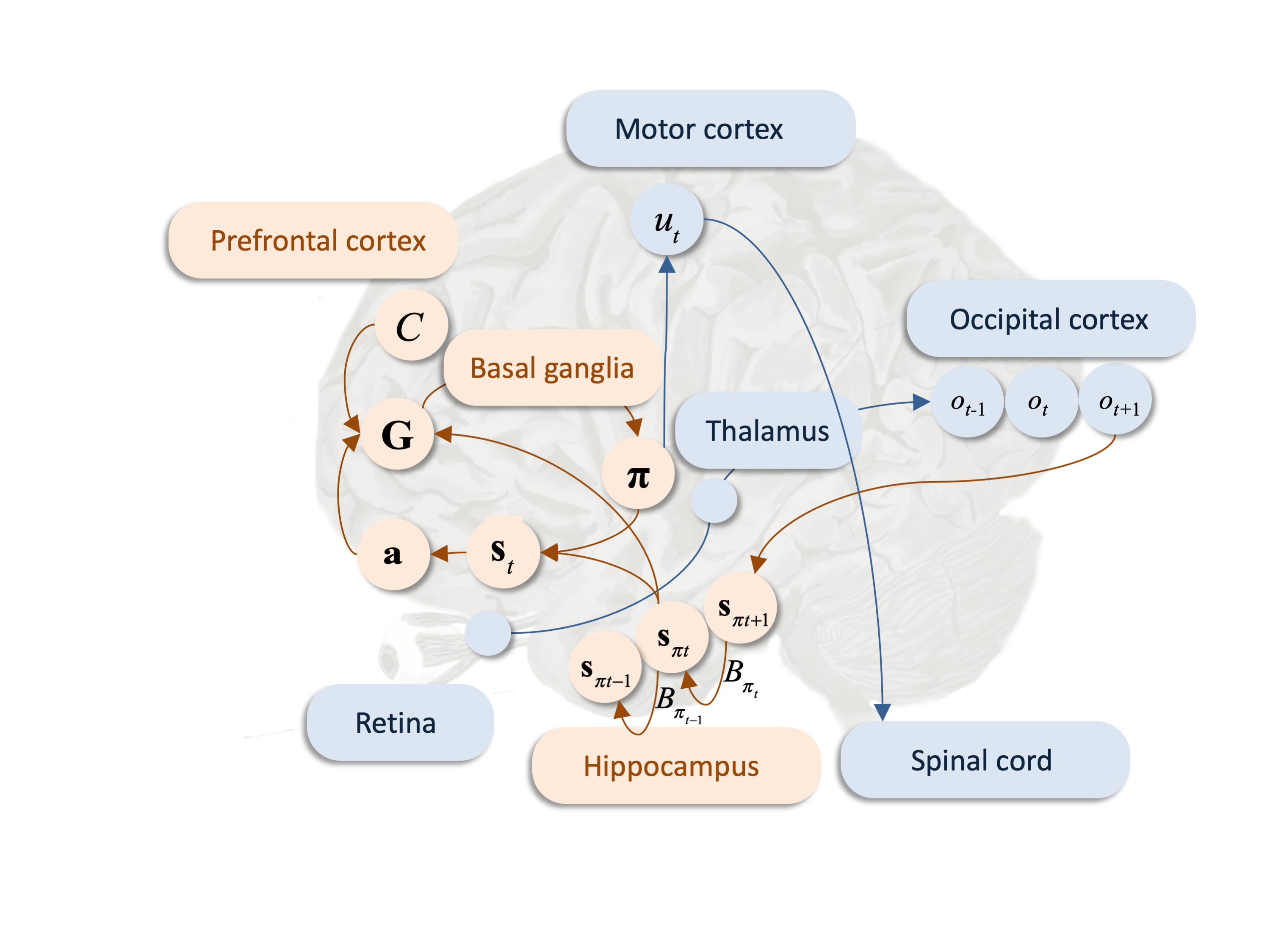}
		\caption{\textbf{Possible functional anatomy.} This figure summarises a possible (coarse-grained) functional anatomy that could implement belief updating in active inference. The arrows correspond to message passing between different neuronal populations. Here, a visual observation is sampled by the retina, aggregated in first-order sensory thalamic nuclei and processed in the occipital (visual) cortex. The green arrows correspond to message passing of sensory information. This signal is then propagated (via the ventral visual pathway) to inferior and medial temporal lobe structures such as the hippocampus; this allows the agent to go from observed outcomes to beliefs about their most likely causes in state-estimation (perception), which is performed locally. The variational free energy is computed in the striatum. The orange arrows encode message passing of beliefs. Preferences $C$ are attributed to the dorsolateral prefrontal cortex – which is thought to encode representations over prolonged temporal scales \cite{parrWorkingMemoryAttention2017} – consistent with the fact that these are likely to be encoded within higher cortical areas \cite{fristonActiveInferenceCuriosity2017}. The expected free energy is computed in the medial prefrontal cortex \cite{fristonActiveInferenceProcess2017} during planning, which leads to inferences about most plausible policies (decision-making) in the basal ganglia, consistent with the fact that the basal ganglia is thought to underwrite planning and decision-making \cite{parrAnatomyInferenceGenerative2018,jahanshahiFrontostriatosubthalamicpallidalNetworkGoaldirected2015,bernsHowBasalGanglia1996,dingBasalGangliaContributions2013,haberPrimateBasalGanglia2003,thibautBasalGangliaPlay2016}. The message concerning policy selection is sent to the motor cortex via thalamocortical loops. The most plausible action, which is selected in the motor cortex is passed on through the spinal cord to trigger a limb movement. Simultaneously, policy independent state-estimation is performed in the ventrolateral prefrontal cortex, which leads to synaptic plasticity dynamics in the prefrontal cortex, where the synaptic weights encode beliefs about $A$.}
		\label{fig: brain}
	\end{figure}

	As one can see, the learning rule concerning accumulation of Dirichlet parameters (c.f., \eqref{eq: learning dirichlet one step}) means that the agent becomes increasingly confident about its likelihood matrix by receiving new observations (since the matrix which is added onto $a$ at each timestep is always positive). This is fine as long as the structure of the environment remains relatively constant. In the next section, we will see how Bayesian model reduction can revert this process, to enable the agent to adapt quickly to a changing environment. Table \ref{table:3} summarises the belief updating entailed by active inference, and Figure \ref{fig: brain} indicates where particular computations might be implemented in the brain.
	
	\begin{longtabu} to \textwidth {
			X[2,c]
			X[4,c]
			X[1,c]}
		\caption{Summary of belief updating.} \label{table:3} \\
		\toprule
		Process & Computation & Equations \\
		\midrule
		{Perception}	&  { $\bold s_{\pi \tau}= \sigma(v), \dot v = - \nabla_{\bold s _{\pi \tau}}F_{\pi} $} &{\eqref{eq:vfe gradients perception}} \\\addlinespace[0.3cm]
		{Planning}	&  {$G(\pi)$ } &{\eqref{eq: sol efe computation}, \eqref{eq: sol efe computation outcomes}} \\\addlinespace[0.3cm]
		{Decision-making}	&  {$Q(\pi)= \sigma(-G(\pi))$} &{\eqref{eq: approx post policies}} \\\addlinespace[0.3cm]
		{Action selection}	&  {$u_t = \arg \max_{u \in U } \left (\sum_{\pi \in \Pi} \delta_{u, \pi_t} Q(\pi)\right)$}&{\eqref{eq: action selection}} \\\addlinespace[0.3cm]
		{Policy-independent state-estimation}	&  {$\bold s_\tau = \sum_{\pi \in \Pi} \bold s_{\pi\tau} Q(\pi)$} &{\eqref{eq: policy indep}} \\\addlinespace[0.3cm]
		{Learning (end of trial)} &  {$\bold a=  a +\sum_{\tau =1}^T o_\tau \otimes \bold s_\tau$} &{\eqref{eq: learning dirichlet one step}} \\ \addlinespace[0.15cm]
		\bottomrule \addlinespace[0.15cm]
	\end{longtabu}
	
	\section{Structure learning}
	\label{sec: structure learning}
	
	In the previous sections, we have addressed how an agent performs inference over different variables at different timescales in a biologically plausible fashion, which we equated to perception, planning and decision-making. In this section, we consider the problem of learning the form or structure of the generative model.
	
	The idea here is that agents are equipped (e.g., born) with an innate generative model that entails fundamental preferences (e.g., essential to survival), which are not updated. For instance, humans are born with prior preferences about their body temperature around $37^\circ$C and $O_2$, $CO_2$, glucose etc concentrations within a certain range. Mathematically, this means that the parameters of these innate prior distributions -- encoding the agent’s expectations as part of its generative model -- have hyperpriors that are infinitely precise (e.g., a Dirac delta distribution) and thus cannot be updated in an experience dependent fashion. The agent’s generative model then naturally evolves by minimising variational free energy to become a good model of the agent’s environment but is still constrained by the survival preferences hardcoded within it. This process of learning the generative model (i.e., the variables and their functional dependencies) is called structure learning.
	
	Structure learning in active inference is an active area of research. Active inference proposes that the agent’s generative model evolves over time to maximise the evidence for its observations. However, a complete set of mechanisms that biological agents use to do so has not yet been laid out. Nevertheless, we use this section to summarise two complementary approaches; namely, Bayesian model reduction and Bayesian model expansion \cite{fristonActiveInferenceCuriosity2017,fristonPostHocBayesian2011,smithActiveInferenceModel2019,fristonBayesianModelReduction2018} – that enable to simplify and complexify the model, respectively.
	
	\subsection{Bayesian model reduction}
	
	To explain the causes of their sensations, agents must compare different hypotheses about how their sensory data are generated – and retain the hypothesis or model that is the most valid in relation to their observations (i.e., has the greatest model evidence). In Bayesian statistics, these processes are called Bayesian model comparison and Bayesian model selection -- these correspond to scoring the evidence for various generative models in relation to available data and selecting the one with the highest evidence \cite{claeskensModelSelectionModel2006,stephanBayesianModelSelection2009}.
	Bayesian model reduction (BMR) is a particular instance of structure learning, which formalises \textit{post-hoc} hypothesis testing to simplify the generative model. This precludes redundant explanations of sensory data – and ensures the model generalises to new data. Technically, it involves estimating the evidence for simpler (reduced) priors over the latent causes and selecting the model with the highest evidence. This process of simplifying the generative model -- by removing certain states or parameters -- has a clear biological interpretation in terms of synaptic decay and switching off certain synaptic connections, which is reminiscent of the synaptic mechanisms of sleep (e.g., REM sleep \cite{hobsonWakingDreamingConsciousness2012,hobsonVirtualRealityConsciousness2014}), reflection and associated machine learning algorithms (e.g., wake-sleep \cite{hintonWakesleepAlgorithmUnsupervised1995}).
	
	To keep things concise, let $\nu$ represent a hidden variable in the generative model that is optimised during learning (e.g. $A$), and $o=o_{1:t}$ a sequence of observations. The current model has a prior $P(\nu)$ and we would like to test whether a reduced prior (i.e., less complex) $\tilde P(\nu)$ can provide a more parsimonious explanation for the observed outcomes. Using Bayes rule, we have the following identities:
	
	\begin{align}
		P(\nu)P(o|\nu)&= P(\nu |o)P(o) \label{eq: bayes 1 structure learning}\\ 
		\tilde P(\nu)P(o|\nu)&= \tilde P(\nu |o)\tilde P(o) \label{eq: bayes 2 structure learning}
	\end{align}
	
	Where $P(o)=\int P(o|\nu)P(\nu) \: d\nu$ and $\tilde P(o)=\int P(o|\nu) \tilde P(\nu)$. Dividing \eqref{eq: bayes 1 structure learning} by \eqref{eq: bayes 2 structure learning} yields
	
	\begin{equation}
		\label{eq: bayes divided}
		\frac{P(\nu)}{\tilde P(\nu)} = \frac{P(\nu|o)P(o)}{\tilde P(\nu|o)\tilde P(o)} 
	\end{equation}
	
	We can then use \eqref{eq: bayes divided} in order to obtain the following relations:
	
	\begin{align}
		1 &= \int \tilde P(\nu|o)\: d\nu =\frac{P(o)}{\tilde P(o)}\int \frac{\tilde P(\nu) P(\nu|o)}{P(\nu)} \:d\nu = \frac{P(o)}{\tilde P(o)}\E_{P(\nu|o)}\left[\frac{\tilde P(\nu)}{P(\nu)}\right] \\
		\Rightarrow &\log \tilde P(o) -\log  P(o) =\log \E_{P(\nu|o)}\left[\frac{\tilde P(\nu)}{P(\nu)}\right] \label{eq: intractable posterir structure learning}
	\end{align}
	
	We can approximate the posterior term in the expectation of \eqref{eq: intractable posterir structure learning} with the corresponding approximate posterior $Q(\nu)$, which simplifies the computation. This allows us to compare the evidence of the two models (reduced and full) and select the best. If the reduced model has more evidence, it implies the current model is too complex – and redundant parameters can be removed by adopting the new priors.
	
	In conclusion, BMR allows for computationally efficient and biologically plausible hypothesis testing, to find simpler explanations for the data at hand. It has been used to emulate sleep and reflection in abstract rule learning \cite{fristonActiveInferenceCuriosity2017}, by simplifying the prior over $A$ at the end of each trial -- this has the additional benefit of preventing the agent from becoming overconfident.
	
	\subsection{Bayesian model expansion}
	
	Bayesian model expansion is complementary to Bayesian model reduction. It entails adopting a more complex generative model (by adding, e.g., more states); if, and only if the gain in accuracy in \eqref{eq: complexity accuracy} is sufficient enough to outweigh the increase in complexity. This model expansion allows for generalisation and concept learning in active inference \cite{smithActiveInferenceModel2019}. Note that additional states need not always lead to a more complex model. It is in principle possible to expand a model in such a way that complexity decreases, as many state estimates might be able to remain close to their priors in place of a small number of estimates moving a lot. This ‘shared work’ by many parameters could lead to a simpler model.
	
	From a computational perspective, concept acquisition can be seen as a type of structure learning \cite{gershmanLearningLatentStructure2010,tervoNeuralImplementationStructure2016} – that can be emulated through Bayesian model comparison. Recent work on concept learning in active inference \cite{smithActiveInferenceModel2019}, shows that a generative model equipped with extra (latent) hidden states can engage these ‘unused’ hidden states, when an agent is presented with novel stimuli during the learning process. Initially the corresponding likelihood mappings (i.e., the corresponding columns of $A$) are uninformative, but these are updated when the agent encounters new observations that cannot be accounted by its current knowledge (e.g., observing a cat when it has only been exposed to birds). This happens naturally, during the learning process, in an unsupervised way through free energy minimization. To allow for effective generalization, this approach can be combined with BMR; in which any new concept can be aggregated with similar concepts, and the associated likelihood mappings can be reset for further concept acquisition, in favour of a simpler model with higher model evidence. This approach can be further extended by updating the number of extra hidden states through a process of Bayesian model comparison.
	
	\section{Discussion}
	Due to the various recent theoretical advances in active inference, it is easy to lose sight of its underlying principle, process theory and practical implementation. We have tried to address this by rehearsing – in a clear and concise way – the assumptions underlying active inference as a principle, the technical details of the process theory for discrete state-space generative models and the biological interpretation of the accompanying neuronal dynamics. It is useful to clarify these results; as a first step to guide towards outstanding theoretical research challenges, a practical guide to implement active inference to simulate experimental behaviour and a pointer towards various predictions that may be tested empirically.
	
	Active inference offers a degree of plausibility as a process theory of brain function. From a theoretical perspective its requisite neuronal dynamics correspond to known empirical phenomena and extend earlier theories like predictive coding \cite{fristonFreeenergyPrincipleUnified2010,raoPredictiveCodingVisual1999,bastosCanonicalMicrocircuitsPredictive2012}. Furthermore, the process theory is consistent with the underlying free energy principle, which biological systems are thought to abide by – namely, the avoidance of surprising states: this can be articulated formally based on fundamental assumptions about biological systems \cite{parrMarkovBlanketsInformation2019,fristonFreeEnergyPrinciple2019}. Lastly, the process theory has a degree of face validity as its predicted electrophysiological responses closely resemble empirical measurements.
	
	However, for a full endorsement of the process theory presented in this paper, rigorous empirical validation of the synthetic electrophysiological responses is needed. To pursue this, one would have to specify the generative model that a biological agent employs for a particular task. This can be done through Bayesian model comparison of alternative generative models with respect to empirical (choice) behaviour being measured (e.g., \cite{mirzaHumanVisualExploration2018}). Once the appropriate generative model is formulated, evidence for a plausible but distinct implementations of active inference would need to be compared, which come from various possible approximations to the free energy \cite{parrNeuronalMessagePassing2019,schwobelActiveInferenceBelief2018,yedidiaConstructingFreeEnergyApproximations2005}, each of which yields different belief updates and simulated electrophysiological responses. Note that the marginal approximation to the free energy currently stands as the most biologically plausible \cite{parrNeuronalMessagePassing2019}. From this, the explanatory power of active inference can be assessed in relation to empirical measurements and contrasted with other existing theories.
	
	This means that the key challenge for active inference -- and arguably data analysis in general -- is finding the generative model that best explains observable data (i.e., evidence maximising). A solution to this problem would enable to find the generative model -- entailed by an agent -- by observing its behaviour. In turn, this would enable one to simulate its belief updating and behaviour accurately \textit{in-silico}. It should be noted that these generative models can be specified manually for the purposes of reproducing simple behaviour (e.g., agents performing simple tasks needed for empirical validation discussed above). However, a generic solution to this problem is necessary to account for complex datasets; in particular, complex behavioural data from agents in a real environment. Moreover, a biologically plausible solution to this problem could correspond to a complete structure learning roadmap; accounting for how biological agents evolve their generative model to account for new observations. Evolution has solved this problem by selecting phenotypes with a good model of their sensory data, therefore, understanding the processes that have selected generative models that are fit for purpose for our environment might lead to important advances in structure learning and data analysis.
	
	Discovering new generative models corresponding to complex behavioural data, will demand to extend the current process theory to these models, in order to provide testable predictions and reproduce the observed behaviour in-silico. Examples of generative models that do not fall within the current discrete state-space, continuous state-space \cite{buckleyFreeEnergyPrinciple2017,fristonActiveInferenceAgency2012,adamsComputationalAnatomyPsychosis2013,brownActiveInferenceSensory2013,adamsPredictionsNotCommands2013,fristonPerceptionsHypothesesSaccades2012,brownFreeEnergyIllusionsCornsweet2012} or mixed \cite{fristonGraphicalBrainBelief2017,parrDiscreteContinuousBrain2018,parrComputationalPharmacologyOculomotion2019} models – currently implemented in active inference – include Markov decision trees \cite{jordanIntroductionVariationalMethods1998,jordanHiddenMarkovDecision1997} and Boltzmann machines \cite{stoneArtificialIntelligenceEngines2019,ackleyLearningAlgorithmBoltzmann1985,salakhutdinovEfficientLearningProcedure2012}.
	
	One challenge that may arise, when scaling active inference to complex models with many degrees of freedom, will be the size of the policy trees in consideration. Although effective and biologically plausible, the current pruning strategy is unlikely to reduce the search space sufficiently to enable tractable inference in such cases. As noted above, the issue of scaling active inference may yield to the first principles of the variational free energy formulation. Specifically, generative models with a high evidence are minimally complex. This suggests that ‘scaling up’, in and of itself, is not the right strategy for reproducing more sophisticated or deep behaviour. A more principled approach would be to explore the right kind of factorisations necessary to explain structured behaviour. A key candidate here are deep temporal or diachronic generative models that have a separation of timescales. This form of factorisation (c.f., mean field approximation) replaces deep decision trees with shallow decision trees that are hierarchically composed.
	
	To summarise, we argue that some important challenges for theoretical neuroscience include finding process theories of brain function that comply with active inference as a principle \cite{parrMarkovBlanketsInformation2019,fristonFreeEnergyPrinciple2019}; namely, the avoidance of surprising events. The outstanding challenge is then to explore and fine grain such process theories, via Bayesian model comparison (e.g., using dynamic causal modelling \cite{fristonDynamicCausalModelling2003,fristonHistoryFutureBayesian2012}) in relation to experimental data. From a structure learning and data analysis perspective, the main challenge is finding the generative model with the greatest evidence in relation to available data. This may be achieved by understanding the processes evolution has selected for creatures with a good model of their environment. Finally, to scale active inference to behaviour with many degrees of freedom, one needs to understand how biological agents effectively search deep policy trees when planning into the future, when many possible policies may be entertained at separable timescales.
	
	\section{Conclusion}
	
	In conclusion, this paper aimed to summarise: the assumptions underlying active inference, the technical details underwriting its process theory, and how the associated neuronal dynamics relate to known biological processes. These processes underwrite action, perception, planning, decision-making, learning and structure learning; which we have illustrated under discrete state-space generative models. We have discussed some important outstanding challenges: from a broad perspective, the challenge for theoretical neuroscience is to develop increasingly fine-grained mechanistic models of brain function that comply with the core tenets of active inference \cite{parrMarkovBlanketsInformation2019,fristonFreeEnergyPrinciple2019}. In regards to the process theory, key challenges relate to experimental validation, understanding how biological organisms evolve their generative model to account for new sensory observations and how they effectively search large policy spaces when planning into the future.
	
	\section*{Software availability}
	The belief updating scheme described in this article is generic and can be implemented using standard routines (e.g., \texttt{spm\_MDP\_VB\_X.m}). These routines are available as Matlab code in the SPM academic software: $\texttt{http://www.fil.ion.ucl.ac.uk/spm/}$. Examples of simulations using discrete state-space generative models can be found via a graphical user interface by typing DEM.
	
	\section*{Acknowledgements}
	
	LD is supported by the Fonds National de la Recherche, Luxembourg (Project code: 13568875). TP is supported by the Rosetrees Trust (Award number: 173346). NS is funded by the Medical Research Council (Ref: 2088828). SV was funded by the Leverhulme Doctoral Training Programme for the Ecological Study of the Brain (DS-2017-026). KF is funded by a Wellcome Trust Principal Research Fellowship (Ref: 088130/Z/09/Z).
	
	\appendix
	
	\section{More complex generative models}
	\label{appendix: more complex models}
	
	In this Appendix, we briefly present cases of more complex discrete state-space generative models and explain how the belief updating can be extended to those cases.
	
	\subsection{Learning B and D}
	\label{appendix:learning B D}
	
	In this paper, we have only considered the case where $A$ is learned, while beliefs about $B$ (i.e., transition probabilities from one state to the next) and $D$ (i.e., beliefs about the initial state) remained fixed. In general, $B$ and $D$ can also be learnt over time. This calls upon a new (extended) expression for the generative model with priors over $B$ and $D$:
	
	\begin{equation}
		\label{eq: gen mod with B and D}
		\begin{split}
			P(o_{1:T},s_{1:T},A,B,D,\pi) &=P(\pi) P(A)P(B)P(D)P(s_1|D)\prod_{\tau=2}^T P(s_\tau|s_{\tau-1},B, \pi) \prod_{\tau=1}^T  P(o_\tau|s_\tau,A)  \\
			P(B)&=\prod_{u\in U}\prod_{i=1}^m P((B_u)_{\bullet i})\quad P((B_u)_{\bullet i})=Dir((b_u)_{\bullet i})\\
			P(D)&=Dir(d)
		\end{split}
	\end{equation}
	Here, $(B_u)_{\bullet i}$ and $(b_u)_{\bullet i}$ denote the i$^{th}$ columns of the matrix $B_u$ encoding the transition probabilities from one state to the next state and its corresponding Dirichlet parameter $b_u$. Furthermore, one needs to define the corresponding approximate posteriors that will be used for learning:
	
	\begin{equation}
		\begin{split}
			Q(B)&= \prod_{u\in U}\prod_{i=1}^m Q((B_u)_{\bullet i})\quad Q((B_u)_{\bullet i})=Dir((\bold b_u)_{\bullet i}) \\
			Q(D)&= Dir(d)
		\end{split}
	\end{equation}
	
	The variational free energy, after having observed $o_{1:t}$, is computed analogously as in equation \eqref{eq: computing VFE}. The process of finding the belief dynamics is then akin to section \ref{sec: learning} -- we rehearse it in the following: selecting only those terms in the variational free energy, which depend on $B$ and $D$ yields:
	
	\begin{equation}
		\begin{split}
			F[Q(B,D)] &= D_{KL}[Q(B)||P(B)] + D_{KL}[Q(D)||P(D)] 
			- \E_{Q(\pi)Q(s_1|\pi)Q(D)}[s_1\cdot \log D] \\ &- \sum_{\tau=2}^t \E_{Q(\pi)Q(s_\tau,s_{\tau-1} |\pi) Q(B)} [s_\tau \cdot \log B_{\pi_\tau}s_{\tau-1}] +\cdots\\
			&= D_{KL}[Q(B)||P(B)] + D_{KL}[Q(D)||P(D)] - \bold s_1 \cdot \textbf{log} \bold D-\sum_{\tau=2}^t \E_{Q(\pi)}[\bold s_{\pi\tau} \cdot \textbf{log} \bold B_{\pi_\tau} \bold s_{\pi_{\tau-1}} ]+ \cdots
		\end{split}
	\end{equation}
	
	Using the form of the KL divergence for Dirichlet distributions \eqref{eq: DKL A} and taking the gradients yields
	
	\begin{align}
		\label{eq: gradient B} \nabla_{\textbf{log} \bold B_u} F(\bold b_u) &=  \bold b_u -b_u -\sum_{\tau =2}^t \sum_{\pi \in \Pi} \delta_{u, \pi_t}  Q(\pi)(\bold s_{\pi\tau} \otimes \bold s_{\pi\tau-1})\\
		\label{eq: gradient D} \nabla_{\textbf{log} \bold D} F(\bold d) &=\bold d -d- \bold s_1
	\end{align}

	where $\otimes$ denotes the Kronecker product. Finally, it is possible to specify neuronal plasticity dynamics following a descent on \eqref{eq: gradient B}, \eqref{eq: gradient D}, which correspond to biological dynamics. Alternatively, we have belief update rules implemented once after each trial of observation epochs in in-silico agents:
	
	\begin{align}
		\bold b_u &= b_u + \sum_{\tau =2}^t \sum_{\pi \in \Pi} \delta_{u, \pi_\tau}  Q(\pi)(\bold s_{\pi\tau} \otimes \bold s_{\pi\tau-1}) \\
		\bold d&= d +\bold s_1
	\end{align}

	\subsection{Complexifying the prior over policies}
	\label{appendix: gamma}
	
	In this paper, we have considered a simple prior approximate posterior over policies; namely, $\sigma (-G(\pi))$. This can be extended to $\sigma (-\gamma G(\pi))$, where $\gamma$ is an (inverse) temperature parameter that denotes the confidence in selecting a particular policy. This extension is quite natural in the sense that $\gamma$ can be interpreted as the postsynaptic response to dopaminergic input \cite{fitzgeraldDopamineRewardLearning2015,fristonAnatomyChoiceDopamine2014}. This correspondence is supported by empirical evidence \cite{schwartenbeckDopaminergicMidbrainEncodes2015} and enables one to simulate biologically plausible dopaminergic discharges (c.f., Appendix E \cite{fristonActiveInferenceProcess2017}). Anatomically, this parameter may be encoded within the substantia nigra, in nigrostriatal dopamine projection neurons \cite{schwartenbeckDopaminergicMidbrainEncodes2015}, which maps well with our proposed functional anatomy (c.f., Figure \ref{fig: brain}), since the substantia nigra is connected with the striatum. We refer the reader to \cite{fristonActiveInferenceProcess2017} for a discussion of the associated belief updating scheme.

	\subsection{Multiple state and outcome modalities}
	\label{appendix: multiple state outcome}
	
	In general, one does not only need one hidden state and outcome factor to represent the environment, but many. Intuitively, this happens in the human brain as we integrate sensory stimuli from our five (or more) distinct senses. Mathematically, we can express this via different streams of hidden states (usually referred to as hidden factors) that evolve independently of one another that interact to generate outcomes at each time step; e.g., see Figure 9 \cite{jordanIntroductionVariationalMethods1998} for a graphical representation of a multi-factorial hidden Markov model. This means that $A$ becomes a multi-dimensional tensor that integrates information about the different hidden factors to cause outcomes. The belief updating is analogous in this case, contingent upon the fact that one assumes a mean-field factorisation of the approximate posterior on the different hidden state factors (see, e.g., \cite{mirzaSceneConstructionVisual2016,fristonFunctionalAnatomyTime2016}). This means that the beliefs about states may be processed in a manner analogous to Figure \ref{fig: brain}, invoking a greater number of neural populations.
	
	\subsection{Deep temporal models}
	\label{appendix: deep temporal models}
	
	A deep temporal model is a generative model with many layers that are nested hierarchically and act at different timescales. These were first introduced within active inference in \cite{fristonActiveInferenceCuriosity2017}. One can picture them graphically as a POMDP (c.f., Figure \ref{fig:gen mod}) at the higher level where each outcome is replaced by a POMDP at the lower level, and so forth.
	
	There is a useful metaphor for understanding the concept underlying deep temporal models: each layer of the model corresponds to the hand of a clock. In a two-layer hierarchical model, a ticking (resp. rotation) of the faster hand corresponds to a time step (resp. trial of observation epochs) at the lower level. At the end of each trial at the lower level, the slower hand ticks once, which corresponds to a time-step at the higher level, and the process unfolds again. One can concisely summarise this by saying that a state at the higher level corresponds to a trial of observation epochs at the lower level. Of course, there is no limit to the number of layers one can stack in a hierarchical model.
	
	To obtain the associated belief updating, one computes free energy at the lower level by conditioning the probability distributions from Bayes rule by the variables from the higher levels. This means that one performs belief updating at the lower levels independently of the higher levels. Then, one computes the variational free energy at the higher levels by treating the lower levels as outcomes. For more details on the specificities of the scheme see \cite{fristonActiveInferenceCuriosity2017}.

	\section{Expected free energy}
	\label{Appendix:steady-state lemma}
	
	At the heart of active inference is a description of a certain class of systems at non-equilibrium steady-state (NESS) \cite{parrMarkovBlanketsInformation2019,fristonFreeEnergyPrinciple2019}. An important consequence of NESS is the existence of a steady-state probability distribution $P(s_\tau,A)$ that the agent is guaranteed to reach given a sufficient amount of time. Intuitively, this distribution should be thought as the agent's preferences over states and model parameters. Practically, this means that the agent selects policies, such that its predicted states $Q(s_\tau,A)$ at some future time point $\tau >t$ -- usually, the time horizon of a policy $T$ -- reach its preferences $P(s_\tau,A)$, which are specified by the generative model. In the following, we will show how a specific family of distributions $Q(\pi)$ guarantee an agent to reach its preferences. Then, we will see how NESS enables in fact to extract one single canonical member of this family: the (softmax negative) \textit{expected free energy}.
	
	\textbf{Objective:} we seek distributions over policies that imply steady-state solutions; i.e., when the final distribution does not depend upon initial observations. Such solutions ensure that, on average, stochastic policies lead to a steady-state or target distribution specified by the generative model. These solutions exist in virtue of conditional independencies, where the hidden states provide a Markov blanket that separates policies from outcomes. In other words, policies cause final states that cause outcomes. In what follows, $\tau >t$ is a future time and $Q:=Q(o_\tau, s_\tau, A, \pi) \approx P(o_\tau, s_\tau, A, \pi|o_{1:t})$ is the corresponding approximate posterior distribution, given initial conditions $o_{1:t}$.
	
	\begin{lemma}[Steady-state]
		\label{lemma:steady-state}
		The surprisal over policies $-\log Q(\pi)$ and the Gibbs energy,
		\begin{align}
			G(\pi; \beta) &= D_{KL}[Q(s_\tau, A|\pi)||P(s_\tau,A)]-\mathbb E_{Q(o_\tau, s_\tau, A|\pi)} [\beta \log P(o_\tau |s_\tau, A)] \\
			\beta &:= \frac{\mathbb E_Q[\log Q(\pi|s_\tau,A)]}{\mathbb E_Q[\log P(o_\tau|s_\tau,A)]} \geq 0
		\end{align}
		are equal on average under $Q$ if and only if the system reaches steady-state. Explicitly:
		\begin{align}
			\mathbb E_Q[-\log Q(\pi)]&= \mathbb E_Q[G(\pi; \beta)] \iff D_{KL}[Q(s_\tau, A) ||P(s_\tau, A)]=0
		\end{align}
	\end{lemma}
	
	Here, $\beta \geq 0$ characterises the steady-state with the relative precision (i.e., negative entropy) of policies and final outcomes, given final states. The generative model stipulates steady-state, in the sense that distribution over final states (and outcomes) does not depend upon initial observations. Here, the generative and predictive distributions simply express the conditional independence between policies and final outcomes, given final states. Note that when $\beta = 1$, Gibbs energy becomes expected free energy.
	
	\begin{proof}
		Let us unpack the Gibbs energy expected under $Q$:
		\begin{equation}
			\begin{split}
				\mathbb E_Q[G(\pi; \beta)] &=\mathbb E_Q[D_{KL}[Q(s_\tau, A|\pi)||P(s_\tau,A)]-\mathbb E_{Q(o_\tau, s_\tau, A|\pi)} [\beta \log P(o_\tau |s_\tau, A)]] \\
				&= \mathbb E_Q[D_{KL}[Q(s_\tau, A|\pi)||P(s_\tau,A)]]\\
				&-\frac{\mathbb E_Q[\log Q(\pi|s_\tau,A)]}{\mathbb E_Q[\log P(o_\tau|s_\tau,A)]} \mathbb E_Q [\mathbb E_{Q(o_\tau, s_\tau, A|\pi)} [ \log P(o_\tau |s_\tau, A)]] \\
				&= \mathbb E_Q[\log Q(s_\tau, A|\pi)-\log P(s_\tau, A)-\log Q(\pi |s_\tau,A)] \\
				&= \mathbb E_Q[-\log Q(\pi)-\log P(s_\tau, A)+\log Q(s_\tau,A)] \\
				&= \mathbb E_Q[-\log Q(\pi)] +D_{KL}[Q(s_\tau,A)||P(s_\tau, A)]
			\end{split}
		\end{equation}
		And the result is immediate.
	\end{proof}
	
	A straightforward consequence of Lemma \ref{lemma:steady-state}, is that each distribution
	\begin{equation}
		\label{eq:posterior_gibbs}
		Q(\pi) =\sigma(-G(\pi; \beta)),\quad \beta \geq 0
	\end{equation}
	
	describes a certain kind of system that self-organises to some steady-state distribution. This family of distributions has interesting interpretations: for example, the case $\beta=0$ corresponds to standard stochastic control, variously known as KL control or risk-sensitive control \cite{vandenbroekRiskSensitivePath2010}: 
	\begin{equation}
		G(\pi; 0) = D_{KL}[Q(s_\tau, A|\pi)||P(s_\tau, A)] \geq D_{KL}[Q(s_\tau |\pi)||P(s_\tau)]
	\end{equation}
	
	In other words, one chooses policies that minimise the KL divergence between the predictive and target distribution. More generally, when $\beta >0$, policies are more likely when they simultaneously minimise the entropy of outcomes, given states. In other words, $\beta >0$ ensures that the system exhibits itinerant behaviour. One can see that KL control may arise in this case if the entropy of the likelihood mapping remains constant with respect to policies.

	\begin{remark}
		It is possible to extend this framework by considering systems that reach their preferences at a collection of time-steps into the future, say $\tau_1,...,\tau_n >t$. In this case, one can adapt the proof of Lemma \ref{lemma:steady-state} to obtain:
		\begin{equation}
			\E_Q \left[\sum_{i =1}^n G(\pi, \tau_i; \beta)\right ]= \E_Q[-n \log Q(\pi)]+ \sum_{i =1}^n D_{KL}[Q(s_{\tau_i},A)||P(s_{\tau_i}, A)]
		\end{equation}
		
		where $G(\pi, \tau_i; \beta)$ is the Gibbs free energy of Lemma \ref{lemma:steady-state}, replacing $\tau$ by $\tau_i$. In this case, the canonical choice of approximate posterior over policies would be:
		
		\begin{equation}
			Q(\pi)= \sigma \left(\frac 1 n \sum_{i =1}^n G(\pi, \tau_i; \beta)\right)
		\end{equation}
		
	\end{remark}
	
	One perspective – on the distinction between simple and general steady-states – is in terms of uncertainty about policies. For example, simple steady-states preclude uncertainty about which policy led to a final state. This would be appropriate for describing classical systems (that follow a unique path of least action), where it would be possible to infer which policy had been pursued, given the initial and final outcomes. Conversely, in general steady-state systems (e.g., mice, Homo sapiens), simply knowing that ‘you are here’ does not tell me ‘how you got here’, even if I knew where you were this morning. Put another way, there are lots of paths or policies open to systems that attain a general steady state.
	
	In active inference, we are interested in a certain class of systems that self-organise to general steady-states; namely, those that move through a large number of probabilistic configurations from their initial state to their final steady-state. The treatments in \cite{parrMarkovBlanketsInformation2019,fristonFreeEnergyPrinciple2019} effectively turn the steady-state lemma on its head by assuming NESS is stipulatively true – and then characterise the ensuing self-organisation in terms of Bayes optimal policies:
	
	\begin{corollary}[Active inference \cite{fristonFreeEnergyPrinciple2019}]
		If a system attains a general steady-state, it will appear to behave in a Bayes optimal fashion – both in terms of optimal Bayesian design (i.e., exploration) and Bayesian decision theory (i.e., exploitation). Crucially, the loss function defining Bayesian risk is the negative log evidence for the generative model entailed by an agent. In short, systems (i.e., agents) that attain general steady-states will look as if they are responding to epistemic affordances \cite{parrWorkingMemoryAttention2017}.
	\end{corollary}
	
	So far, we have deduced the distribution over policies of systems that reach steady-state. However, recall that reaching steady-state is only a consequence of NESS. In fact, NESS dynamics under a Markov blanket (c.f., Figure \ref{fig: markov blanket}) imply a slightly stronger statement: the most likely trajectories of systems described by active inference are those which minimise \textit{expected free energy} \cite{parrMarkovBlanketsInformation2019,fristonFreeEnergyPrinciple2019} -- this is exactly the case $\beta =1$ in \eqref{eq:posterior_gibbs}. This is nice, since many existing theories of cognition and control emerge under this specific imperative (c.f., Figure \ref{fig: EFE interpretations}).
	
	\section{Computing expected free energy}
	
	In this appendix, we present the derivations underlying the analytical expression of the expected free energy that is used in \texttt{spm\_MDP\_VB\_X.m}. Following \cite{parrComputationalNeurologyActive2019}, we can reexpress the expected free energy in the following form:
	
	\begin{multline}
		\label{eq: G new expression}
		G(\pi)= \underbrace{\mathbb E_{Q(s_\tau|\pi)} [\text{H}[P(o_\tau |s_\tau)]]}_{\text{Ambiguity}} + \underbrace{D_{KL}[Q(s_\tau |\pi)||P(s_\tau)]}_{\text{Risk (states)}}  
		- \underbrace{\mathbb E_{P(o_\tau |s_\tau)Q(s_\tau|\pi)}[D_{KL}[Q(A|o_\tau,s_\tau)||Q(A)]]}_{\text{Novelty}}
	\end{multline}
	
	Here, $Q(A|o_\tau,s_\tau)$ denotes approximate posterior beliefs about $A$ if we knew occurence of the state outcome pair $(o_\tau,s_\tau)$. In the following, we show that we can compute the expected free energy in the following way
	\begin{equation}
		\label{eq: sol efe computation}
		\begin{split}
			G(\pi) &\approx \underbrace{H \cdot \bold s_{\pi \tau}}_{\text{Ambiguity}} + \underbrace{ \bold s_{\pi \tau} \cdot (\log \bold s_{\pi \tau}-\log C)}_{\text{Risk (states)}} - \underbrace{\bold A \bold s_{\pi \tau} \cdot W \bold s_{\pi \tau}}_{\text{Novelty}} \\
			H &:= - \text{diag} [\bold A\cdot \log \bold A] \\
			W &:= \frac 1 2 \left(\bold a ^{\odot (-1)}-\bold a_0^{\odot (-1)}\right)
		\end{split}
	\end{equation}
	
	when the agent's preferences $C$ are expressed in terms of preferences over states. When preferences are expressed in terms of outcomes (as is currently implemented in \texttt{spm\_MDP\_VB\_X.m}), the risk term instead becomes
	
	\begin{equation}
		\label{eq: sol efe computation outcomes}
		\underbrace{(\bold A \bold s_{\pi\tau}) \cdot ( \log (\bold A \bold s_{\pi\tau}) - \log C)}_{\text{Risk (outcomes)}}
	\end{equation}
	
	\subsection{Ambiguity}
	
	The ambiguity term of \eqref{eq: G new expression} is $\mathbb E_{Q(s_\tau|\pi)} [\text{H}[P(o_\tau |s_\tau)]]$. By definition, the entropy inside the expectation is:
	
	\begin{equation}
		\text{H}[P(o_\tau |s_\tau )]=-\sum_{o_\tau \in O} P(o_\tau |s_\tau )\log P(o_\tau|s_\tau ) 
	\end{equation}
	
	The first factor inside the sum corresponds to:
	
	\begin{equation}
		\begin{split}
			P(o_\tau |s_\tau) &=\int P(o_\tau,A|s_\tau)\: dA \\
			&=\int P(o_\tau |s_\tau,A)P(A)\:dA \\
			&=\int o_\tau \cdot As_\tau P(A)\:dA \\
			&\approx \int o_\tau \cdot As_\tau Q(A) \: dA \\
			&= o_\tau \cdot \mathbb E_{ Q ( A )} [A] s_\tau  \\
			&=o_\tau \cdot \bold As_\tau 
		\end{split}
	\end{equation}
	
	Here we have replaced the prior over model parameters $P( A)$ by the approximate posterior $Q( A)$. This is not necessary, but in numerical simulations since learning occurs once at the end of the trial the two can be interchanged -- furthermore, this allows us to reuse previously introduced notation. In any case, this tells us that the entropy can be re-expressed as:
	
	\begin{equation}
		\begin{split}
			\text{H}[P(o_\tau |s_\tau )] &= - \sum_{o_\tau \in O} (o_\tau \cdot \bold As_\tau )(o_\tau \cdot \log (\bold A)s_\tau ) \\
			&= - \sum_{i=1}^n (\bold A_{\bullet i}s_\tau)(\log(\bold A_{\bullet i})s_\tau) \\
			&= - \sum_{i=1}^n (\bold A_{\bullet i}\odot \log (\bold A_{\bullet i}))s_\tau \\
			&= - (\bold A\odot \log \bold A)s_\tau\\
			&= - \text{diag} [\bold A \cdot \log \bold A ] \cdot s_\tau 
		\end{split}
	\end{equation}
	
	Finally,
	
	\begin{equation}
		\begin{split}
			\underbrace{\mathbb E_{Q(s_\tau|\pi)} [\text{H}[P(o_\tau |s_\tau)]]}_{\text{Ambiguity}} &= H\cdot \bold s_{\pi \tau} \\
			H := - \text{diag} [\bold A \cdot & \log \bold A ]
		\end{split}
	\end{equation}

	\subsection{Risk}
	
	The risk term of \eqref{eq: G new expression} is the KL divergence between predicted states following a particular policy and preferred states. This can be expressed as:
	
	\begin{equation}
		\underbrace{D_{KL}[Q(s_\tau|\pi)||P(s_\tau)]}_{\text{Risk (states)}}= \bold s_{\pi\tau} \cdot (\bold s_{\pi\tau} - \log C)
	\end{equation}
	
	Where the vector $C \in \mathbb R^m$ encodes preference over states $P(s_\tau)=Cat(C)$. However, it is also possible to approximate this risk term over states by a risk term over outcomes (c.f., \eqref{eq:pref outcomes}), as is currently implemented in \texttt{spm\_MDP\_VB\_X.m}. In this case, if $C \in \mathbb R^n$ denotes the preferences over outcomes $P(o_\tau)=Cat(C)$:
	
	\begin{equation}
		\underbrace{D_{KL}[Q(o_\tau|\pi)||P(o_\tau)]}_{\text{Risk (outcomes)}} =(\bold A \bold s_{\pi\tau}) \cdot ( \log (\bold A \bold s_{\pi\tau}) - \log C)
	\end{equation}

	\subsection{Novelty}
	
	The novelty term of \eqref{eq: G new expression} is ${\mathbb E_{P(o_\tau|s_\tau)Q(s_\tau|\pi)}[D_{KL}[Q(A|o_\tau,s_\tau)||Q(A)]]}$ where
	
	\begin{align}
		Q(A) &= \prod_{i=1}^m Q(A_{\bullet i}), \quad Q(A_{\bullet i}) = Dir(\bold a_{\bullet i}) \\
		Q(A|o_\tau, s_\tau) &= \prod_{i=1}^m Q(A_{\bullet i}|o_\tau, s_\tau), \quad Q(A_{\bullet i}|o_\tau, s_\tau) := Dir(\bold a'_{\bullet i}) 
	\end{align}
	
	The KL divergence between both distributions (c.f., \eqref{eq: DKL A}) can be expressed as:
	
	\begin{equation}
		\label{dkl novelty}
		\begin{split}
			D_{KL}[Q(A|o_\tau, s_\tau) || Q(A)] =\sum_{i=1}^m[\log \Gamma(\bold a '_{0 i}) -\sum_{k=1}^n\log \Gamma(\bold a '_{k i})-\log \Gamma(\bold a _{0 i})\\
			+\sum_{k=1}^n\log \Gamma(\bold a_{k i})]+(\bold a ' - \bold a)\cdot (\psi(\bold a ')-\psi(\bold a '_0)) 
		\end{split}
	\end{equation}

	where $\psi$ is the digamma function. We now want to make sense of $\bold a'$. Suppose that at time $\tau$ the agents knows the possible outcome $j$ and possible state $k$ as in $Q(A|o_\tau,s_\tau)$ (c.f., Table \ref{table:2} for terminology). This means that in this case, beliefs about hidden states correspond to the true state; in other words, $\bold s_\tau =s_\tau$. We can then use the rule of accumulation of Dirichlet parameters to deduce $\bold a'=\bold a +o_\tau \otimes s_\tau $. In other words, $\bold a'_{jk} = \bold a_{jk}+1$ and the remaining components are identical. Using the well-known identity:
	
	\begin{equation}
		\Gamma(x+1)= x\Gamma(x)\Rightarrow \log \Gamma(x+1) = \log x + \log \Gamma(x)
	\end{equation}
	
	we can compute \eqref{dkl novelty}:
	
	\begin{equation}
		\begin{split}
			D_{KL}[Q(A|o_\tau, s_\tau) || Q(A)] &= \log \Gamma(\bold a_{0k}+1) -\log \Gamma(\bold a_{0k})-\log \Gamma (\bold a_{jk}+1)+ \log \Gamma (\bold a_{jk})+\psi(\bold a_{jk}+1)-\psi(\bold a_{0k} \\
			&=\log \bold a_{0k}-\log \bold a_{jk}+\psi (\bold a_{jk}+1)-\psi (\bold a_{0k}+1)
		\end{split}
	\end{equation}
	
	Using the definition of the digamma function $\psi(x)=\frac{d}{dx} \log \Gamma(x)$ we obtain:
	
	\begin{equation}
		\begin{split}
			D_{KL}[Q(A|o_\tau, s_\tau) || Q(A)] &= \log \bold a_{0k}-\log \bold a_{jk} +\frac{d}{d \bold a_{jk}}(\log \Gamma(\bold a_{jk}+1))- \frac{d}{d \bold a_{0k}}(\log \Gamma(\bold a _{0k}+1))\\
			&= \log \bold a_{0k}-\log \bold a_{jk} +\frac{d}{d \bold a_{jk}}(\log \Gamma(\bold a_{jk}+1))- \frac{d}{d \bold a_{0k}}(\log \bold a _{0k} +\log \Gamma(\bold a _{0k})) \\
			&= \log \bold a_{0k}-\log \bold a_{jk} + \frac{1}{\bold a_{jk}}-\frac{1}{\bold a_{0k}} +\psi(\bold a_{jk})-\psi(\bold a_{0k})
		\end{split}
	\end{equation}
	
	We can use an asymptotic expansion of the digamma function to simplify the expression:
	
	\begin{equation}
		\begin{split}
			\psi(x) &\approx \log x -\frac{1}{2x}+\cdots \\
			\Rightarrow D_{KL}[Q(A|o_\tau, s_\tau) || Q(A)] &\approx \frac{1}{2\bold a_{jk}}-\frac{1}{2\bold a_{0k}}
		\end{split}
	\end{equation}
	
	Finally, the analytical expression of the novelty term:
	
	\begin{equation}
		\begin{split}
			\E_{P(o_\tau |s_\tau)Q(s_\tau|\pi)}[D_{KL}[Q(A|o_\tau, s_\tau) || Q(A)]] &\approx \bold A \bold s_{\pi \tau}\cdot  W \bold s_{\pi\tau}\\ 
			W :=\frac{1}{2}\left( \bold a^{\odot -1}-\bold a_0^{\odot -1} \right)&
		\end{split}
	\end{equation}

	%\printbibliography
	\bibliographystyle{unsrt}
	\bibliography{aiexplained_bib}

\end{document}